\documentclass[journal]{new-aiaa}
\usepackage[utf8]{inputenc}
\usepackage{textcomp}
\usepackage{graphicx}
\usepackage{amsmath}
\usepackage{longtable,tabularx}
\usepackage[table, xcdraw]{xcolor}
\usepackage{multirow}
\usepackage{rotating}
\usepackage{float}
\title{Voyager 3: A Concept Mission to Interstellar Medium}

\author{Shohreh Abdolrahimi\footnote{Assistant Professor, Physics and Astronomy, 3801 W Temple Ave Pomona CA 91768.}, Burton Yale\footnote{Undergraduate Student, Aerospace Engineering, 3801 W Temple Ave Pomona CA 91768, Student Member.}, Christos C. Tzounis\footnote{Adjunct Professor, Physics and Astronomy, 3801 W Temple Ave Pomona CA 91768.}, Joshua Fofrich$^\dagger$, Rohan Patel$^\dagger$, Jehosafat Cabrera-Guzman$^\dagger$, Jonathan C. Welsher$^\dagger$, and Navid Nakhjiri\footnote{Associate Professor, Aerospace Engineering, 3801 W Temple Ave Pomona CA 91768, Member.}}
\affil{California State Polytechnic University, Pomona, California, 91768}
\author{David Scott\footnote{Enterprise System Engineer, Jet Propulsion Laboratory, 4800 Oak Grove Drive, M/S 301-460, Pasadena, CA  91109}, and April Johnson\footnote{Institutional Process Specialist, Jet Propulsion Laboratory, 4800 Oak Grove Drive, M/S 301-460, Pasadena, CA  91109}}
\affil{Jet Propulsion Laboratory, California Institute of Technology, Pasadena, California, 91109}

\begin{document}

\maketitle

\begin{abstract}
The Voyager 3 is a concept mission that sends a space telescope to the interstellar medium in a reasonable amount of time. Voyager 3 would take a direct image of an exoplanet using the solar gravitational lensing at the distance of 550 AU. The spacecraft would use its suite of scientific instruments to study the local solar system's environment, interstellar medium and finalize its primary mission by imaging an exoplanet. Two potential architectures are proposed to meet this mission directive, using multiple gravitational assists and sizable electric propulsion burns to achieve the high escape speeds necessary to reach 550 AU. This paper expands on the science mission objectives, trajectory, and the preliminary design of the baseline spacecraft. 
\end{abstract}

%


\section{Introduction}
\lettrine{M}{ore} than 40 years ago, in 1977, Voyager 1 and Voyager 2 began an ambitious mission to explore the outer solar system and gather observations directly at the source, from outer planets we had only seen with remote studies. The spacecrafts were built to accomplish their planetary missions and last for five years. However, both spacecraft are still sending scientific information about their surroundings through the Deep Space Network. With the Voyager 1 and 2 reaching the end of their operational lifetime, and other spacecraft such as the New Horizons continuing the exploration of the Interstellar Medium (ISM), deep-space missions are at the forefront of expanding our knowledge of space outside of our solar system. The search for life beyond our planet has always been an exciting endeavor, wrought with infinite possibilities, only limited by our own technology.  Voyager 3 mission would follow the path of its predecessors Voyager 1 and 2 to explore the interstellar medium (ISM) and make a direct image of an exoplanet using solar gravitational lensing (SGL). 
   
On Aug. 25, 1989, NASA's Voyager 2 spacecraft made the first and also the last close flyby of Neptune, completing the end of the Voyager mission's grand tour of the solar system. Eventually, between them, Voyager 1 and 2 explored all the giant outer planets of our solar system and 48 of their moons revolutionizing the science of planetary astronomy. Recently, they both embarked on the scientific exploration of interstellar space. On August 2012, Voyager 1 made its historic entry into interstellar space. Voyager 2 entered interstellar space on November 5, 2018. The two Voyager spacecrafts have now confirmed that the plasma in local interstellar space is significantly denser than the plasma inside the heliosphere. 
The data from Voyager 2 has allowed the direct analysis of the cosmic frontier for the first time \cite{P24}. The results show that the plasma barrier is significantly hotter and thicker than previous studies estimated, effectively forming a physical shield between our solar system and interstellar space stopping about $70\%$ of cosmic radiation from breaking into our solar system. Voyager 2 observations show that the temperature is $30,000-50,000~\text{K}$, whereas models and observations predicted a very local interstellar medium (VLISM) temperature of $15,000-30,000~\text{K}$. Voyager 3 would further explore the interstellar medium, investigating the extragalactic background light and dust/plasma. 

Within the last few decades, over 4,200 exoplanets have been discovered. Each prompted discovery has spurred a desire to know more about these exoplanets. In the effort to search for life on other planets, the main hindrance has always been how fine of detail can be resolved. In order to investigate the possible hospitality of a planet, the first aspect that must be considered is the atmospheric composition, elements such as $\textrm{O}_2$, $\textrm{CO}_2$, and $\textrm{N}_2$. Current imaging capabilities of these objects are nowhere near the level required for accurate atmosphere spectroscopy. To solve this discrepancy, we propose a space mission architecture to utilize the concept of Solar Gravitational Lensing (SGL). By using the Sun's considerable gravitational attraction, a telescope would be able to increase its imaging resolution by orders of magnitude, allowing for the near megapixel image quality of a planet around another star. Much like glass lenses, the observation must happen from a focal point. For the Sun's case, this focal point is a line running from $550$ AU to $1,000$ AU. To get a telescope to these distances, innovative designs must be considered, and this is why the mission is following in the legacy of the previous Voyager missions. This paper aims to provide the preliminary design of such a vehicle to accomplish this goal.

SGL has gained prominence in recent years with the release of the NIAC report by Turyshev~\cite{TuryshevReport}. In the report, Turyshev proposed several mission architectures that ranged from traditional to experimental. The major measuring criteria for the mission type is the time of flight (ToF) and exit velocity. The shorter the ToF the more reliable the hardware on the spacecraft can be expected to be. One such architecture proposed would use multiple small spacecraft with solar sail technology for the main source of propulsion~\cite{TuryshevReport}. Through the use of interferometeretry, the group of spacecrafts would combine their large array of smaller instruments, to make an effectively larger telescope. The spacecraft would launch from Earth and perform a flyby of Jupiter to bring the heliocentric periapsis close to the sun. The close proximity to the Sun allows for a two-headed approach to increasing the speed of the spacecraft. As the solar sail nears the Sun, its effective thrust would increase by $1/r^2$, this in addition to the Oberth effect allows the spacecraft to further increase its energy gaining potential~\cite{TuryshevReport}. The current limiting factor on the technology is its inability to produce solar sails light enough and strong enough for a spacecraft with adequate science payloads~\cite{TuryshevReport}.

Another common architecture for this mission could be a singular large spacecraft using a more traditional mode of propulsion, such as chemical and/or electric thrusters. The architecture would use a single spacecraft  that holds a large $1m$ diameter telescope. The spacecraft would have a chemical propulsion system and, like the previous architecture, perform an Oberth Maneuver at the Sun~\cite{ret15}. Alternatively, another mission design possibility would be to have the spacecraft perform a similar maneuver at Jupiter~\cite{Reza}. Trajectories that utilize a chemical propulsion powered flyby would allow the spacecraft to leave the solar system at velocities of up to 14 AU/year~\cite{ret15}. 

Assuming that the architecture would use a single spacecraft with a large +1m diameter telescope, nuclear electric propulsion (NEP) is a potential candidate for the propulsion type. NEP provides the opportunity to gain large amounts of \(\Delta V\) over time and has been considered in the past~\cite{Reza}. One potential mission design would use NEP combined with traditional planetary flybys to escape the solar system. This method would provide a robust trajectory that can leave the solar system {and reach the SGL focal point} with flight times of less than 60 years. A hybrid solution utilizing both chemical propulsion and NEP has also shown to have ToFs of less than 50 years~\cite{Reza}. The spacecraft would use an Oberth maneuver, at either Jupiter or the Sun, to leave the solar system and then use electric propulsion to gain a higher exit velocity than with the powered flyby alone. {Reducing the mission time below 50 years would require technologies at a much lower readiness level, such as low-perihelion solar sail trajectories~\cite{TURYSHEV2018}.}

{Of the above concepts, a combination of NEP and a high-energy flyby of Jupiter was chosen for having the highest average readiness level technologies.} Voyager 3 would be the first spacecraft able to directly view a planet around another star. To accomplish this tasks, it would travel 3.9 times further than Voyager 1 is currently, to 550 astronomical units (AU). Along the way to the SGL focal point, the spacecraft would use its suite of scientific instruments to study the environment of the local solar system and beyond. The Voyager 3 spacecraft is proposed, primarily, as a telescoping mission with an additional set of instruments and labs onboard to support various scientific objectives, which will be discussed in this paper. Similar to the previous Voyager missions, Voyager 3 aims to raise the bar in terms of new flying equipment, particularly in the fields of nuclear spacecraft propulsion, as it would be necessary to complete the mission within a reasonable time frame. 

In this paper, we first introduce the need for science and explore { many} mission objectives that can be addressed by this mission. Furthermore, the paper introduces a primary architecture that can achieve this mission and deliver the science data back to Earth. An alternative architecture is also discussed at the end if a solar flyby and science measurements in extended ranges need to be considered. {Among the science goals introduced in this paper, our primary goal and our science baseline is the SGL, reaching 550AU and successfully imaging the surface of the exoplanet, at least 1 km$^2$ of the planet surface. Voyager 3 investigates science objective identified by the Planetary Science Decadal Survey, for the details refer to appendix A and Table 5. A science traceability matrix (STM) is presented in Table~\ref{tab:scienceTraceMatrix}. Table~\ref{tab:measurementObjectives} summarizes each science objective and provides the science instruments needed for proposed measurements on the spacecraft.  }


\section{Science Mission} \label{sec:science}

Voyager 3 provides various opportunities for science experiments. The mission is designed for delivering data over a long period of time and stay active beyond the Solar focal point to transfer the data back to Earth. Considering that the spacecraft’s distance to the sun would range from less than 1 AU to more than 550 AU, it can create a profile for changes of the solar activities in an extended range beyond any other spacecraft. 

\noindent For Voyager 3 three main categories of science objectives (SO) are considered: 
\begin{enumerate}
\item Phase I: Solar System Science (SO1)
\item Phase II: Interstellar Medium Science (SO2)
\item Phase III: Imaging an exoplanet (SO3)
\end{enumerate} 

In this section, we address each of the science drivers and questions from Phase I, II, and III of the Voyager 3 mission, and how they would benefit the scientific community. The overall goal of SO1 is to understand how the magnetic field, solar winds, and Zodiacal light (dust cloud) vary throughout the solar system, including the outer regions, and searching for the possible violations of the weak equivalence principle. The overall goal of SO2 is to investigate the Sun's birth environment and provide and new data for the expansion rate of the universe. The overall goal of SO3 is to investigate the properties of exoplanets using a solar gravitational lens. Table~\ref{tableSQ} provides an overall summary of the Science Objectives and the driving question behind each one. These objectives are further discussed in this section.

\begin{table}[hbt!]
\caption{\label{tableSQ} Science Questions}
\centering
\begin{tabular}{p{1.5in} p{4.5in}}
 \hline
 \multicolumn{2}{c}{Phase I Science} \\
 \hline
Jupiter Flyby (SO1.1)& \begin{itemize}\item How did the giant planet's atmosphere form and evolve to its present state?
\item What is the current impact rate on Jupiter and to what extent can its composition be used as a record of its impact history?
\item What are the characteristics of bodies and large airbursts and how do these compare to known bodies and airbursts to that of Earth? \end{itemize}    \\ \hline
 Solar Magnetic Fields (SO1.2)& \begin{itemize} \item Understand the distribution of magnetic fields and matter throughout the solar system \item How does the interplanetary field interact with the galactic field? \end{itemize}\\ \hline
 Solar Winds (SO1.3)&\begin{itemize} \item How does the solar wind vary throughout the solar system? \item What is the nature of the termination shock and heliosheath? \item What are the properties of the heliopause transition region? \item How does the solar wind interact with the ISM? \end{itemize}\\ \hline
Zodiacal Dust cloud (SO1.4)&\begin{itemize} \item What were the initial stages, conditions, and processes of solar system formation? \item What is the nature of interstellar matter that was incorporated? \item How does the zodiacal dust cloud serve as models for dust clouds of other extrasolar systems? \end{itemize}\\ \hline
 Extragalactic Background Light (EBL) (SO1.5)&\begin{itemize} \item What were the first objects to light up the universe, and when did they do it?
 \end{itemize}\\ \hline
Small Body Science (SO1.6)&\begin{itemize} \item What were the initial stages, conditions, and processes of solar system formation and the nature of the incorporated interstellar matter? \item How have the myriad chemical and physical processes that shaped the solar system operated, interacted, and evolved? \end{itemize}\\
 \hline
\end{tabular}

\begin{tabular}{p{1.5in} p{4.5in}}
 \hline
 \multicolumn{2}{c}{Phase II Science} \\
 \hline
Interplanetary Dust and Plasma (SO2.1)& \begin{itemize}\item What was it about the Sun's birth environment or its star formation process that determined the final properties of our solar system versus that of other planetary systems? \item How much gas and dust were left over for planet formation in our solar system? \end{itemize}    \\ \hline
  Parallax Science (SO2.2)&\begin{itemize} \item At what rate is the universe expanding? \end{itemize}\\
\hline
\end{tabular}

\begin{tabular}{p{1.5in}p{4.5in}}
 \hline
 \multicolumn{2}{c}{Phase III Science} \\
 \hline
 Solar Gravitational Lensing (SGL) (SO3.1)&\begin{itemize} \item Does life exist on exoplanets?
\item What are the properties of exoplanets?  \end{itemize}\\
\hline
\end{tabular}

\end{table}

\subsection{Phase I Science Objectives}
The first set of objectives for the Voyager 3 mission are related to understanding the solar system and create a better model for phenomena that impact our understanding of planets, moons, and even beyond our solar system (also inspired by~\cite{P22}). 
 
\subsubsection{Jupiter Flyby (SO1.1)}
In 1973 and 1979 Pioneer 10 and Voyager 1 and 2 were the first three spacecraft to visit the gas giant. Voyager 1 and 2 discovered Jupiter's faint rings and several new moons and volcanic activity on Io's surface. Between 1995 and 2003, the Galileo spacecraft dropped a probe into Jupiter's atmosphere and conducted extended observations of Jupiter and its moons and rings. In 2000, Cassini took a highly detailed true-color mosaic photo of the gas giant. 
New Horizons passed Jupiter on Feb. 28, 2007, while it used the planet's gravity assist to increase its speed, it also tested its instruments in flight. Although it was the eighth spacecraft to visit Jupiter and its moons, it made discoveries. Accordingly, it recorded lightning, shedding new perspectives on Jupiter's atmospheric storms, recording aurorae in Jupiter's atmosphere, volcanic eruptions on the moon Io, the life cycle of fresh ammonia cloud, boulder-sized clumps speeding through the planet's faint rings, and the pulsing of Jupiter's magnetosphere. Since 2016, Juno spacecraft has arrived at Jupiter, conducting an in-depth investigation of the planet's atmosphere, deep structure, and magnetosphere analysis for clues to its origin and evolution. The goal of Voyager 3 mission at Jupiter is to follow up on the previous and current missions (especially the Juno mission) during its flyby of this planet.

\subsubsection{Solar Magnetic Fields (SO1.2)}
The heliosphere is often defined as a ``bubble-like'' region of space surrounding the sun and planetary system that arose due to the solar wind; more importantly, it is one the main reasons life exists within our solar system as it shields the planets from harmful galactic cosmic radiation. However, there are several features of the heliosphere that are still not very well understood, most stemming from the debates concerning the structure of the heliosphere itself~\cite{P22}. The magnetic field is one of the key elements for better understanding the structure of the heliosphere and the physical processes that take place within it; thus, the first science objective of Phase I of the Voyager 3 mission examines these magnetic field lines to determine how they influence the overall shape and structure of the heliosphere. The complex fundamental physical processes that govern how energy and matter are transported within our solar system, can also play a role in other astrophysical settings. In that sense, the Sun, the heliosphere, and Earth's magnetosphere and ionosphere serve as cosmic laboratories for studying universal plasma phenomena, with applications to laboratory plasma physics, fusion research, and plasma astrophysics. The last mission to measure the magnetic fields directly via in-situ observations was Voyager 1 and 2. Measurements from Voyager 2 confirmed the asymmetric structure of the heliosphere, due to the titled interstellar magnetic field~\cite{P23}. Another observation from Voyager was that magnetic flux in the heliosheath is not conserved, with the reason still unknown~\cite{P24}. While the Voyagers were able to obtain measurements of the magnetic fields as they traveled throughout the solar system, the equipment was not sensitive enough for the regions of space where the magnetic fields were weak. Voyager 3 is the perfect opportunity to follow up on these measurements and survey the unknown field ahead in the ISM. Here we should note that SO1.2 addresses key science goals 3 and 4 from NASA Decadal Strategy for Solar and Space Physics~\cite{P39}.

 \subsubsection{Solar Winds (SO1.3)}
The outer heliosphere is divided into different regions: the termination shock, heliosheath, and heliopause. In front of the heliosphere, a bow shock is believed to form as the solar system travels through the ISM, although the presence is still being debated \cite{P22}. Another important component is the pick-up ion (PUI) population, which are solar wind protons that interact with the interstellar hydrogen atoms. In fact, $80\%$ of the energy in termination shock is dominated by these PUIs, which were not measured by Voyager, leading to an energy gap in the data. While this energy gap may also be due to energetic electrons, only a revisit to this region with proper instrumentation helps to fill this energy gap \cite{P25}. Some of the ongoing debates concerning the nature and structure of the heliosphere include the degree to which the heliosphere influences the local ISM, how the ISM magnetic field drapes around the heliosphere, the physical phenomena responsible for the observed ``tailless'' two-lobe shape of the HP, and other arguments summarized by the works of Ed Stone, et al. \cite{P22}. Many of these conflicts can only be resolved by performing more in-situ measurements of the magnetic fields that would be possible along the trajectory of Voyager 3. Here we should note that SO1.3, addresses key science goals 3 and 4 from NASA Decadal Strategy for Solar and Space Physics \cite{P39}.

 \subsubsection{Zodiacal Dust Cloud (SO1.4)}
The overall goal of the Zodiacal Dust Cloud study is to understand the initial stages, conditions, and processes of solar system formation, and the nature of interstellar matter that was incorporated through the investigation of the zodiacal light, which would consequently result in the understanding of the dust distribution within the solar system. The current model of the interplanetary dust cloud (IPDC)~\cite{P26} assumes a smooth cloud, with 3 dust bands, and a circumsolar dust ring. Recent observations suggest the presence of another circumsolar dust ring near Mercury orbit~\cite{P27}. Near-Earth observations of zodiacal light are unable to produce a unique physical model of IPD and observations from outside the IPDC are necessary. Moreover, it would serve as a model for the dust and the zodiacal light pollution for the exoplanetary systems. Modeling the zodiacal light in the exoplanetary systems is motivated by the observation of the Very Large Telescopic Interferometer (VLTI), which has discovered that the exozodiacal light (zodiacal light around other star systems) poses a challenging obstacle to the direct imaging of Earth-like planets. 

\subsubsection{Extragalactic Background Light (SO1.5)}
The goal of Extragalactic Background Light part of the mission is to address the question of: ``What were the first objects to light up the universe, and when did they do it?'' by making a precise measurement of the extragalactic background light (EBL). The EBL spectrum captures cosmological backgrounds associated with photons emitted by stars, galaxies, and active galactic nuclei (AGN) due to nucleosynthesis or other radiative processes, including dust scattering, absorption and reradiation or primordial phenomena, such as the cosmic microwave background (CMB), at microns to mm-wave radio wavelengths. The EBL may also contain signals that are diffuse and extended, including high energy photons associated with dark matter particle decays or annihilation~\cite{P28}. The precise measurement of EBL could be used to constrain models of galaxy formation and evolution.  With a precise measurement of the EBL intensity spectrum, a window into new discoveries could be opened with profound implications for astronomy ranging from recombination signatures during reionization and diffuse photons associated with dark matter annihilation and their products (see review in~\cite{P29}). A precise direct measurement of the cosmic optical background (COB) can be achieved using a small aperture telescope observing at distances of 5AU or more~\cite{P29}. 

Due to the expansion of the universe, light emitted from the creation and evolution of early galactic structures is expected to be redshifted into the near-IR and IR wavelengths~\cite{Hauser2001}. Therefore, it is of great interest to obtain accurate measurements of the extragalactic background light (EBL) in these wavelengths in order to probe the history of early sources as well as the nature of early galactic evolution such as star formation rates and initial mass functions. The objective of the Voyager 3 mission is to directly measure the EBL intensity at 1.75, 2.5, and 5.3$\mu$m, the nominal wavelengths of the HgCdTe detectors that would be used on the spacecraft. Direct measurements are difficult and not yet conclusive, owing to the large uncertainties caused by the bright foreground emission associated with zodiacal light. The constraints on the cosmic optical background (COB) obtained from an analysis of the Pioneer 10/11 Imaging Photopolarimeter (IPP) data with accurate starlight subtraction was achieved by referring to all-sky star catalogs and a Galactic stellar population synthesis model down to 32.0 mag \cite{Matsuoka2011}.  

Within the range of wavelengths we are interested in, there have been calculations of the EBL using measurements from DIRBE and subtracting the stellar flux from galactic stars using the 2MASS All-Sky Point Source Catalogue that set the intensity of the EBL at 14.69$\pm$4.49kJysr$^{-1}$ at 2.2 $\mu$m, 15.62$\pm$3.34kJysr$^{-1}$ at 2.5 $\mu$m, and 8.88$\pm$6.26kJysr$^{-1}$ at 1.25 $\mu$m~\cite{Levenson2007}. Because the EBL measurements quoted here are residuals after subtracting stellar flux and foreground light, there is little room for improvement without another large sky survey, which Voyager 3 is not capable of. The best measurements of COB come from an indirect technique involving $\gamma$-ray spectra of bright blazars with an absorption feature resulting from pair-production off of COB photons. COB has large uncertainties involving direct measurements due to uncertain removal of the zodiacal light foreground. At optical and near-IR wavelengths, the EBL intensity is predominantly due to stellar emission from nucleosynthesis throughout cosmic history (see~\cite{Hauser} for a review). The COB spectrum also includes radiative information from the reionization epoch. The dominant limitation for direct EBL intensity spectrum at these wavelengths is the zodiacal light associated with scattered solar light from micrometer-size interplanetary dust (IPD) particles near the Earth's orbit. Current measurements with DIRBE on COBE make use of a model to remove zodiacal light or slight variations~\cite{97,27,P29}.
 
The Zodiacal light and its pollution in our inner solar system have long been a challenge for the study of the early Universe for the cosmologists. There is much to be gained from simply performing the same observations from outside of the interplanetary dust cloud, where measurements aren't dominated by the zodiacal light leading to large uncertainties. For this reason, a team at NASA, JPL and Caltech has been looking into the possibility of hitching an optical telescope to a survey spacecraft on a mission to the outer solar system, such as in the ZEBRA concept mission~\cite{ZEBRA}. Voyager 3, investigates the EBL and takes on further missions. 

We set the requirement that Voyager 3 must be able to measure the intensity of the EBL at or below 8.0 kJysr$^{-1}$, and expect a significant reduction in uncertainty by completely eliminating the contribution from the IPD cloud which was previously estimated using models and found to contribute up to $90\%$ of the total light emitted in the wavelength regime of interest~\cite{Kelsall:1998bq}. 
 
\subsubsection{Small body (SO1.6)}
Studies of primitive bodies encompass asteroids, comets, Kuiper belt objects (KBOs), the moons of Mars, and samples, meteorites and IPD particles, derived from them~\cite{P1}. Primitive bodies are thought to have formed earlier than the planets in the hierarchical assembly of solar system bodies, have witnessed, or participated in, many of the formative processes in the early solar nebula. Therefore, these objects provide unique information on the solar system's origin and early history and help researchers to interpret observations of debris disks around other stars~\cite{P1}. There are more than 100,000 KBOs accounted for in this region and many more yet to be discovered beyond the orbit of Neptune (30AU) to approximately 50AU.
The study of primitive bodies over the past decade has been accomplished as a result of several space missions such as Deep Impact, Stardust, EPOXI, Cassini, and the Japan Aerospace Exploration Agency's (JAXA's). Hayabusa has provided a sample of a near-Earth asteroid; determining that small asteroids can be rubble piles~\cite{P1}. Ground-and space-based telescopes and radar studies have discovered that binary objects are common among near-Earth, main-belt asteroids, and KBOs~\cite{P1}. Cassini had come to the first encounter with a possible former trans-Neptune object in the form of Phoebe, Saturn's satellite. Deep Space has determined the density of a comet nucleus via the first controlled cratering experiment on a primitive body. However, important questions remain unanswered~\cite{P1}. 
Necessary steps have been accomplished in constraining the nature and mixing during the formation of primitive bodies, as well as formation chronology~\cite{P15, P16}. Diverse components in comet samples returned by the Stardust spacecraft and the composition of the Sun constrained by the detailed analysis of the solar wind samples from Genesis spacecraft illustrate a dynamical nebular mixing~\cite{P18}.  However, the size distributions of KBOs, now moderately well known for bodies greater than 100 km, are serving as the best tracer of the accretion phase in the outer solar nebula~\cite{P19}. In addition, the structure of the Kuiper belt provides one of the best constraints on the dynamical rearrangement of the giant planets, and some recent KBO studies have revised scenarios for the early orbital history of the solar system~\cite{P20}. Ground-based telescopes continue the discovery and characterization of a significant portion of KBOs. The 3-meter NASA Infrared Telescope Facility (IRTF) has provided significant data for studies of primitive bodies. The Large Synoptic Survey Telescope (LSST) project was a top-rated priority for ground-based telescopes for the years 2011-2021 by the 2010 Astronomy and Astrophysics Decadal Survey~\cite{P21}. Even though ground-based telescopes continue to be relevant to the study of larger or closer objects, Voyager 3 would provide a unique close view and extended observation time to KBOs.

\subsection{Phase II Science Objectives}
 
\subsubsection{Interplanetary Dust \& Plasma (SO2.1)}
The study of Interplanetary Dust \& Plasma has the goal to understand what it was about the Sun's birth environment or its star formation process that determined the final properties of our solar system versus that of other planetary systems; in addition, Voyager 3 would attempt to determine how much gas and dust was left over for the planet formation in our solar system. Dust grains are a very significant component for the understanding of the heating and cooling of the ISM. Moreover, dust is also important in the study of the interstellar chemistry which is essential in the formation of most organic and inorganic molecules~\cite{P33, P34, P22}. The Voyagers 1 and 2, and the New Horizons missions investigated interplanetary dust and plasma. However, a follow-up study with Voyagers 3 new instruments would address many of the yet open questions as mentioned in Table~\ref{tableSQ}. 
 
\subsubsection{Parallax Science (SO2.2)}
The goal of Parallax Science is to address the question: ``At what rate is the universe expanding?''  Dark energy is hypothesized to explain the tending to accelerate the expansion of the universe. As stated in the Astronomy and Astrophysics Decadal Survey (2010)~\cite{P21}, the only way moving forward in understanding this mysterious component of our universe is to use the universe at large to infer the properties of dark energy by measuring its effects on the expansion rate and the growth of structure, (addressing the Frontier of Knowledge, Expansion of the Universe). Therefore, precision measurements of the expansion of the universe with time and of the rate at which cosmic structure grows are required. There are two distinct methods for the measurement of the Hubble constant, which gives us the current rate of the expansion of the universe, and  these methods provide two very different values for the constant. 
The discrepancy between the results of these two methods implies that there could be new physics underlying the foundations of the universe. Possibilities include the interaction strength of dark matter, dark energy being even more exotic than previously thought, or an unknown new particle in the tapestry of space~\cite{P36}. With the second data release of the Gaia mission, the parallax measurement of hundreds of Cepheid variables has been determined with a precision on the order of $10\mu as$~\cite{P37}, allowing the most precise measurement of $H_{0}$ to date. However, Gaia is still limited by its {measurement} baseline distance of $\sim$1AU. The Voyager 3 can extend the {measurement} baseline to 550 AU. {Although the actual measurement quality could be worse than Gaia}. 

\subsection{Phase 3 Science Objectives}
 
\subsubsection{Solar Gravitational Lensing (SO3.1)}
According to the general theory of relativity, rays of light passing in the vicinity of a massive object are deflected from their initial direction by the amount of $\theta=4GM/c^2b$, where $G$ is the gravitational constant, $M$ is the mass of the object, $c$ is the speed of light, and $b$ is the rays impact parameter. This phenomenon is called gravitational lensing.  Gravitational lensing is a well-known effect and has been observed where relatively nearby galaxies, or clusters of galaxies, act as gravitational lenses for background galaxies, and in our Galaxy where micro-lensing of stars in the Galactic bulge or in the Magellanic clouds are caused by intervening bodies. Astronomers have used it to measure the shape of stars, look for exoplanets, measure dark matter in distant galaxies, and measure the size and age of the Universe. In our solar system, the effect was originally observed by Eddington in 1919 \cite{Eddington}. The gravitational field of the sun acts as a spherical lens to magnify the intensity of radiation from a distant source along a semi-infinite focal line, creating a powerful instrument, solar gravitational lens (SGL). Depending on the impact parameter of the incoming ray, electromagnetic (EM) waves traveling from distant sources in the close proximity of the Sun are focused by the solar gravitational field at heliocentric distances beyond $b^2/(2r_g)\leq 547.6 (b/R_\odot)^2$AU, where $b$ is a light ray's impact parameter, $r_g=2GM_\odot/c^2$ is the Schwarzschild radius of the Sun and $R_\odot$ is its radius. The focus of the SGL starts at $\sim$547 AU and going beyond 2,500 AU~\cite{Kraus, Turyshev}, forming a focal line. One advantage of the gravitational lens is the amplification of the power received at the telescope. A mission to the gravitational focus of the Sun was listed as a precursor goal in the 1998 Jet
Propulsion Laboratory (JPL) workshop “Robotic Interstellar
Exploration in the Next Century”~\cite{R7}. A mission utilizing SGL could provide direct multi-pixel
high-resolution images and spectroscopy of a potentially habitable
Earth-like exoplanet. 

To construct an image using the solar gravity lens, Voyager 3 must travel to approximately 550AU and maintain a precise alignment along the focal line of the Sun's gravity lens. Due to the precise alignment, Voyager 3 would not be used to search for habitable exoplanets; the extreme magnification of the solar gravity lens means that we would simply be unable to cover enough sky space to reliably locate exoplanets. Rather, Voyager 3 focuses on the characterization of a previously discovered potentially habitable exoplanet, selected before the mission is launched. It is unlikely that the trajectory of the spacecraft allows for imaging of more than one system, so the target must be chosen carefully. Factors affecting the target choice include host star spectral type, distance to the host star, and the number of exoplanets around the host star, and the type of exoplanets around the star. The ideal candidate would be a nearby, Sun-like star hosting multiple rocky exoplanets in or near the star's habitable zone.

At the focal line, light from an exoplanet forms an annulus, surrounding the edge of the Sun, which is much dimmer than the Sun. We refer to this as the Einstein ring. The remarkable optical properties of SGL include major brightness
amplification ($8\pi^2GM/(c^2\lambda)$, $\sim 10^{11}$ at $\lambda=1\mu$m, ) and extreme angular resolution of $0.1 (\lambda/1\mu \text{m})(650 \text{AU}/r_o)^{1/2}$ nas~\cite{Turyshev2020}. The apparent brightness of sun at 550 AU is $\sim 4.5\times 10^{-3} \text{W}/\text{m}^2$. However, even with this amount of amplification, the apparent brightness of the planet is $1.85\times 10^{-10} \text{W}/\text{m}^2$, for a planet which is 12 parsec from Sun. When averaged over a 1-m telescope, the light amplification is reduced. However, this is still sufficiently bright, even on the solar corona background. {\text{red}The contribution of the solar corona is removable, A preliminary coronagraph design was analyzed in~\cite{TuryshevReport}. The concept here used for the imaging of the exoplanet is to use a single telescope, but the light from different parts of the Einstein ring is imaged separately. Then, the signals from the different parts of the ring are then deconvoluted using an algorithm to reconstruct the planet~\cite{Landis}. It is also important to note that before the SGL mission is launched, one has to gather information about the target exoplanet and its system, as much as possible. The planet's orbit would have to be measured using astrometry or radial velocity measurements. Any other information such as the orbital ephemeris, estimates of the rotation rate, and some understanding of cloud and surface properties from Doppler imaging, could pave the way for the further information that would be gained from SGL.}

The solar gravitational lens addresses the Astronomy and Astrophysics decadal survey science frontier discovery area of the identification and characterization of nearby habitable exoplanets''. There has been a surge of exoplanet discoveries, especially in the Earth and super-Earth mass regime, which was previously not well-detected. NASA's Transiting Exoplanet Survey Satellite mission and the future James Webb Space Telescope will lead to thousands more detected exoplanets, many of which could be potentially habitable. Detection of these potentially habitable exoplanets is an issue that is currently being addressed by many parties across the globe; however, characterization of habitable exoplanets is extremely difficult and limited to planets which transit their host star and planets which we can directly image. While direct imaging has been successful for some systems, it is limited to nearby systems with very giant planets due to the optical limits of our current telescopes. Transit spectroscopy has been more successful, and we have been able to characterize the atmospheres of their transit spectra. Unfortunately, transit spectroscopy requires both proper orbital alignment and very precise observation times, making it difficult or impossible for many exoplanet systems. Additionally, while there are many global atmospheric models from which we can make predictions about the planetary climate based on atmospheric parameters, the models are not definitive, and we cannot predict surface conditions with certainty. The resolution of the solar gravitational lens would allow for spatially resolved images of the surface of the exoplanet, addressing the questions of habitability such as the presence of liquid water or vegetation.

\subsection{Measurements Objectives}
Voyager 3 investigates each science objective by a series of measurements listed in Table~\ref{tab:measurementObjectives}. The table also provides the science instruments needed for proposed measurements on the spacecraft. The measurement objectives are referring to the technology needed for each SO and map the instrument necessary for each objective. In addition, the measurement objectives reflect the mission goal to advance knowledge by increasing accuracy and decreasing uncertainty on science experiments that have been carried out by other missions.
 
Our approach includes combining a set of 7 instruments on the spacecraft such that all science objectives can be measured and addressed with a combination of not more than a few instruments. The list of measurements demonstrates our power management philosophy such that the initial phases of the mission, may require 3-4 instruments to operate for an experiment while the final phases of SO2 and SO3 only rely on 1-2 instruments. Refer to appendix for the science traceability matrix and other tables, i.e.,~\ref{tab:measurementObjectives}.

\section{Mission Architecture} \label{sec:primaryarch}

To meet this mission directive, many possible architecture types were considered, including solar sails, constellation interferometry, and low Earth orbit spacecraft construction, but in terms of reliability and technical promise, the possible mission architectures were narrowed down to two options. The primary configuration utilizes multiple gravitational assists and large electric propulsion burns out to achieve the high escape speeds necessary to reach 550 AU. The alternative architecture leverages the Sun's gravity well to increase the efficiency of a large chemical burn, near the Sun's surface. In the following sections, each of the primary architecture's subsystems will be defined. Then, the technical differences of the alternative architecture will be discussed.

\subsection{Mission Requirements}
To view the topology of an exoplanet, first, an architecture must be defined that can reach 550 AU, and beyond, reliably, while carrying the required equipment. As such, this spacecraft shall be able to reach 550 AU in under 60 years, to have enough power to transmit back all its scientific data. Additionally, the spacecraft shall have an attitude control system capable of performing the fine detail maneuvers to view the Einstein ring. Finally, the architecture shall carry a telescope instrument with the main mirror diameter at or larger than 1 meter. This size telescope would allow the spacecraft to meet the scientific objectives defined in Section~\ref{sec:science}.

\subsection{Science Instruments and Payload}
The payload instruments for the spacecraft were selected to meet the previously mentioned science and measurement objectives. As the main goal of the mission is to examine the activity of an exoplanet, the large diameter telescope is the primary instrument for this mission. The ARIEL telescope, or derivative of, was chosen for the spectrum band its imager covers (\( 0.5 - 7.8 \mu m \))~\cite{ret12}, as well for its lightweight configuration and power requirements. All additional instruments are listed in Table~\ref{tab:retInstruments}, including their associated science objective. {While this mission includes a wide range of science objectives, the major focus of Voyager 3 is to provide updated information on the heliopause using modern equipment (SO2), and more critically the exoplanet science (SO3.1). It is worth emphasizing that Voyager 3 can perform better the objectives SO3.1, SO2.2, SO1.5 than most other architectures. Though we note that there are near-Earth missions such as HabEx that address SO3.1. For many of the rest of objectives we don’t have to wait 60 years for any science return.  }

\begin{table}[htb]
  \caption{Science instrument payload for primary architecture}
  \label{tab:retInstruments}
  \centering
  \begin{tabular}{c c c c c}
    \hline 
    \textbf{Instrument}            & \textbf{Science Objective \#} & \textbf{ Purpose}                & \textbf{Mass (kg)} & \textbf{Power (W)} \\ \hline
    Wide Angle Camera~\cite{JunoCam}              & SO1.1, SO1.6                & Small Body Observation       & 3.7                 & 5.9                 \\ \hline     
    MAG--3~\cite{ret9}                         & SO1.2                          & Magnetic Field Study         & 0.1                & 0.03               \\ \hline                
    SWICS~\cite{ret10}                          & SO1.3, SO2.1                  & Ion Composition {Spectroscopy} & 6                  & 6                  \\ \hline                 
    Cosmic Dust Analyzer~\cite{ret11}           & SO1.4, SO2.1                  & Interstellar Dust {Analysis}   & 16                 & 11                 \\ \hline  
    PEPSSI~\cite{Pepssi}                        & SO1.1                         & Plasma Density {Measurement}   & 1.5                 & 2.5                 \\ \hline                
    ARIEL~\cite{ret12}                          & SO1.5, SO2.2, SO3.1           & Exoplanet Imaging            & 180                & 140                \\ \hline                 
  \end{tabular}
\end{table}

\subsection{Concept of Operations}
\begin{figure}[htb]
  \centering
  \includegraphics[width = \textwidth]{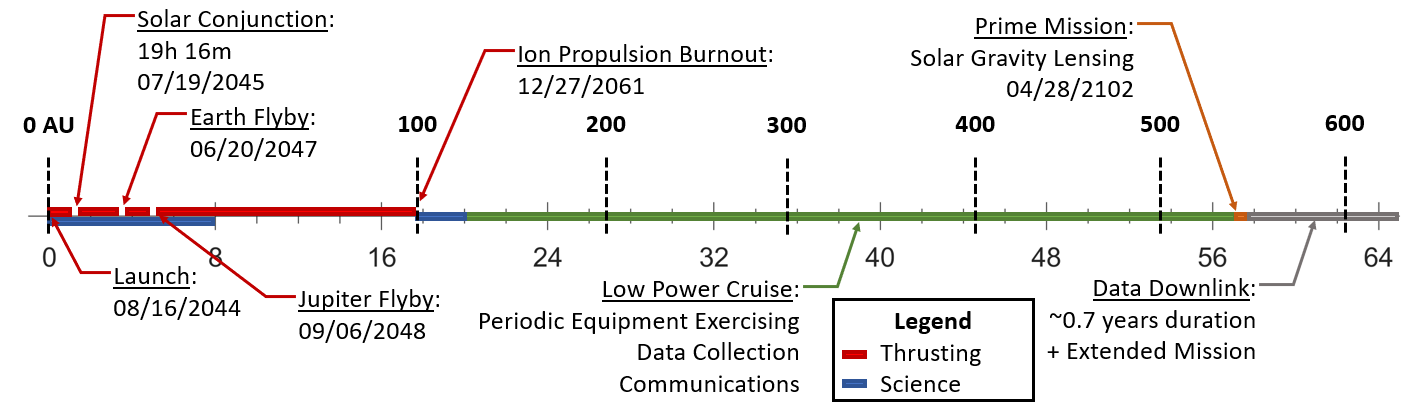}
  \caption{Mission timeline show major mission milestones and science opportunities, with tick marks for years since launch, and radial distance from Sun in AU}
  \label{fig:retTimeline}
\end{figure}

The Voyager 3 mission would begin in August of 2044 with a Space Launch System (SLS) Block 1B~\cite{ret14} launch and heliocentric orbital injection. After a systems checkout period, the spacecraft begins a low thrust burn to increase its heliocentric energy in preparation for an additional earth flyby, as in Figure~\ref{fig:retTimeline}. During the interplanetary orbit and Earth gravity assist, which occurs in 2047, the spacecraft can conduct science and calibrate its onboard instruments. After another checkout period, the low-thrust engines restarts and heads towards Jupiter in 2048 for the final gravity assist, concluding the collection of Phase I science. A long-duration low-thrust burn, depleting all remaining 8600 kg of propellant, would send Voyager 3 into the interstellar medium at its final velocity of 52.6 km/s. At this point, the spacecraft enters its deep-cruise phase of the mission. 

Phase II science would be conducted in addition to periodic equipment exercising and communication. During this period, the telescope, and accompanying equipment, would also be available to the scientific community for outside experiments. This availability would extend from the completion of the heliopause, T+20 years after launch to the crossing of the solar focal line, T+58 years. Once Voyager 3 reaches this point it would begin the imaging of the TRAPPIST-1 exoplanet system. The data would be sent to Earth, and the possible extended mission and disposal protocol begins. Figure~\ref{fig:retTimeline} summarizes the critical phases of the mission in the order they would occur.

\subsection{Trajectory Design}
The primary goal of the trajectory design is to be able to target and fly past a specific point in space directly opposite of the TRAPPIST-1 star system. This point would have a fixed radial distance from the Sun of 550 AU. It would be oriented in space with a specific right ascension of 166.62 degrees and a declination angle of 5.04 degrees with respect to the Mean Ecliptic J2000 reference frame. 

\begin{figure}[htb]
  \centering
  \includegraphics[width = \textwidth]{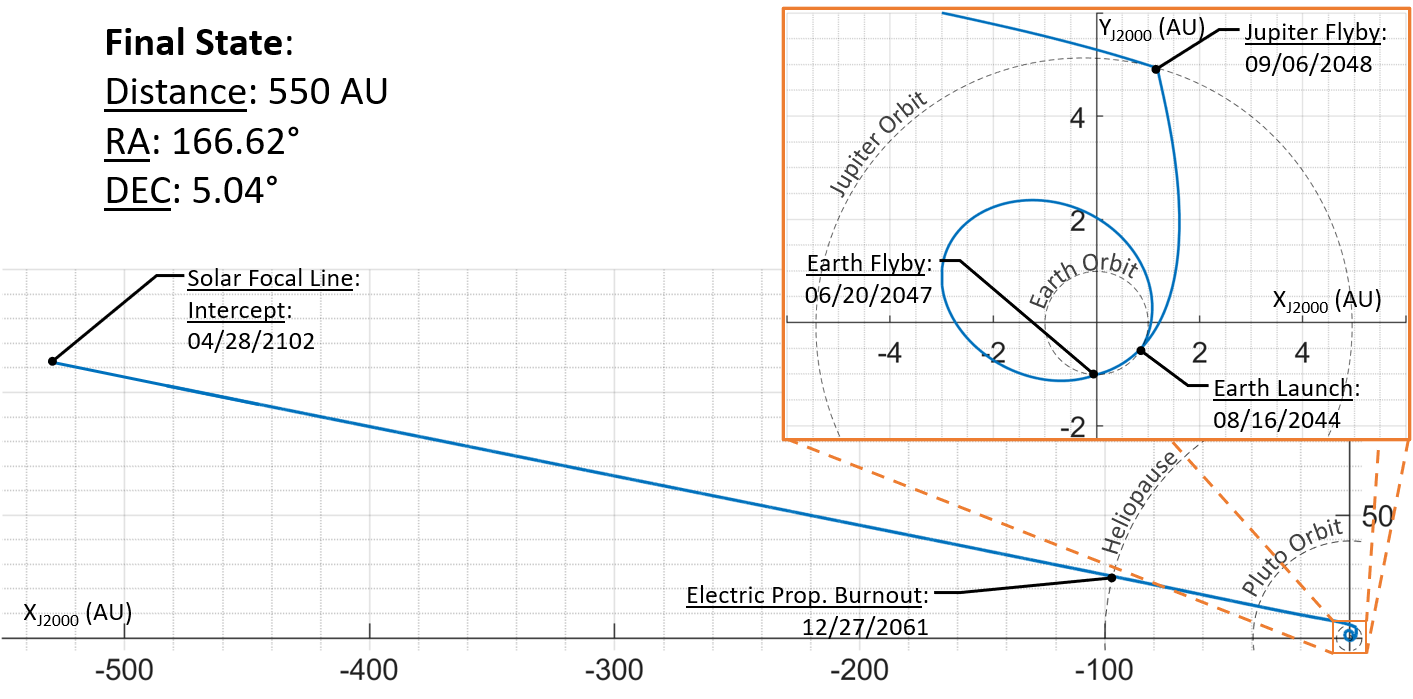}
  \caption{Full overview of Voyager 3's trajectory from the interplanetary flyby's to ion propulsion burn out to arrival at solar focal line.}
  \label{fig:retTrajOverview}
\end{figure}

Another key objective of the trajectory design is to maximize the solar system heliocentric velocity to reach the solar focal line in a reasonable time. Planetary gravity assists can help boost the spacecraft’s velocity post flyby, but still, a significant amount more \(\Delta V\) is required. Therefore, a low-thrust, long-duration burn trajectory would be considered for this mission design. By combining classical interplanetary trajectory design with Hall Effect engine { technology}, a reasonable time of flight was achieved.

A multi-gravity assist, broad trajectory search was conducted before the optimization process to find possible flybys that would maximize the solar system escape velocity. This data was provided by the Jet Propulsion Laboratory's broad search algorithm named Star.~\cite{ret17}

\begin{figure}[htb]
  \centering
  \includegraphics[width = 0.75\textwidth]{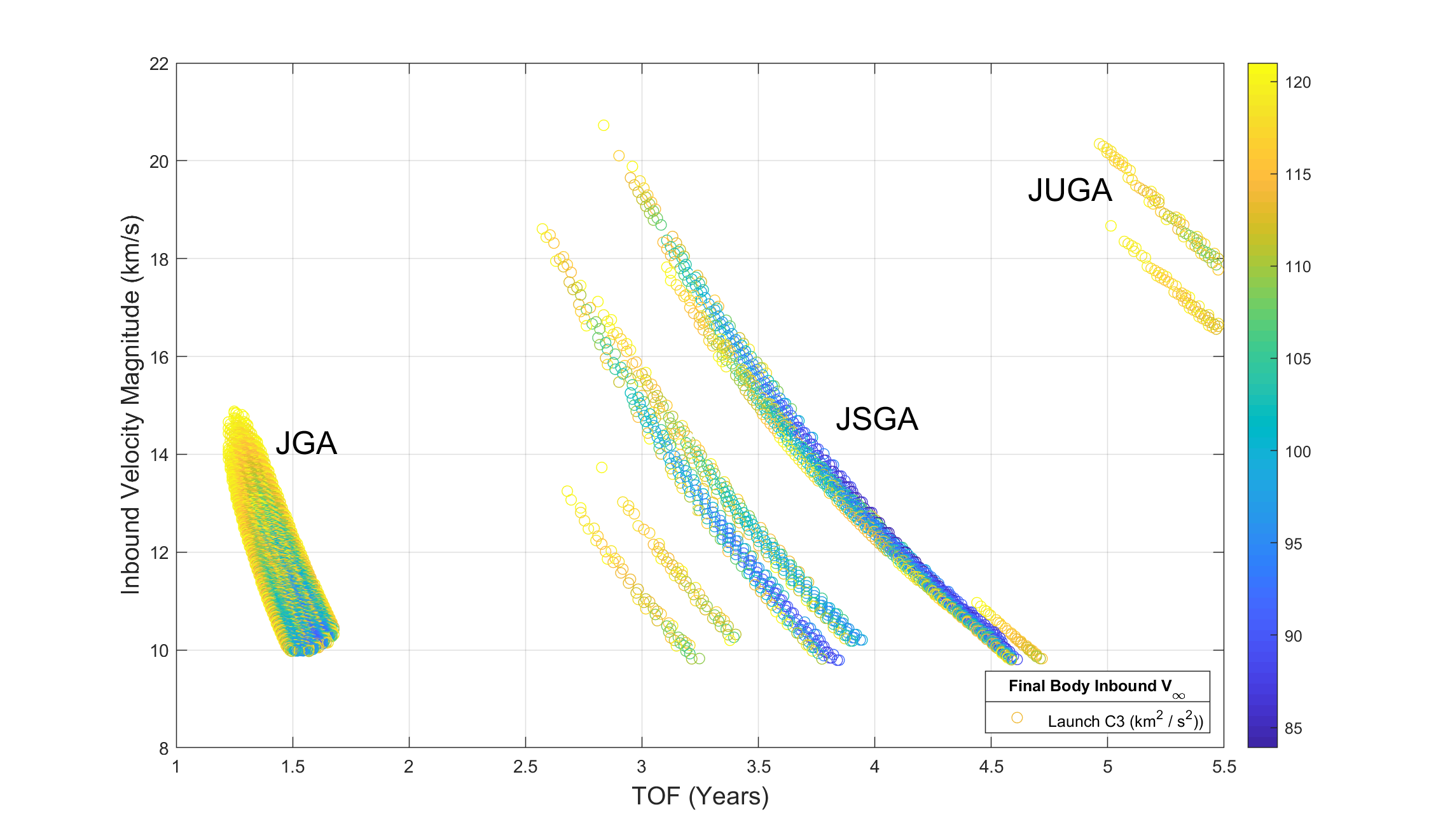}
  \caption{Broad search results of possible trajectory sequences to 550 AU, color graded by launch C3. While JUGAs provide the highest inbound velocity magnitude, the lead up time to the spacecrafts arrival eliminates all possible gains.}
  \label{fig:retStar}
\end{figure}

Figure~\ref{fig:retStar} plots the resulting trajectory possibilities by their time of flight, inbound relative velocity, and the required Earth departure characteristic energy. The search yielded several sets of gravity assist combinations which were traded based on their inbound relative velocity (\(V_\infty\)), right ascension of the final body at the encounter, maximum achievable turn angle (assuming a planar flyby and close approach altitude of no less than 5000 km), and the encounter year. The Saturn only gravity assist trajectories were immediately dropped for their {high} Earth departure C3 and their low \(V_\infty\) values. In the search, three different candidates were analyzed for preliminary investigation: a Jupiter, Gravitational Assist (JGA) with the Jupiter encounter in January of 2048, a JSGA with the Saturn encounter in November of 2059, and a JUGA with the Uranus encounter in 2039. The JSGA and JUGA trajectories offer higher inbound \(V_\infty\) values of 20.73 and 20.35 km/s, respectively, compared to the 14.03 km/s from a single Jupiter flyby. The additional gas giant flyby trajectories provide the opportunity for more Phase I planetary science, along with their increased speed. However, there are certain limitations in these flyby body candidates. Both GA combinations had post flyby right ascension discontinuities, with respect to the target, of greater than 11 degrees. This meant that the low thrust propulsion system would have to change the direction of travel and boost the heliocentric velocity of the spacecraft, resulting in a slower burnout velocity. Also, a more aggressive design and manufacturing timeline would be necessary for the JUGA trajectory as it encounters Uranus in 2039. Thus, the Earth launch would have to be in the late 2020s to accommodate for \(\Delta V\) leveraging flybys and the transfer to Jupiter. Subsequently, the JSGA opportunity encounters Saturn, which is roughly 9.6 AU from the Sun, in 2059. With these considerations and the fact that the launch period becomes much more sensitive to the multi-flyby mission design, a Jupiter only gravity assist trajectory was selected for optimization. To finalize trajectory elements and generate ephemerides, JPL's Mission Analysis Low Thrust Optimizer (MALTO) is utilized~\cite{MALTO}.

The broad search found the Earth to Jupiter (EJ) portion of the trajectory at the maximum C3 (120 \(\text{km}^2 / \text{s}^2\)). {The initial phase of the trajectory} would be possible on current, conventional launch vehicles, but the {mass these vehicles can deliver to such transfers is too small to carry Voyager 3}. The initial estimate for Voyager 3 was on the order of 2500 kg dry and in excess of 12,000 kg wet. On a direct transfer to Jupiter, the SLS Block 1B would only be able to send 5000 kg into orbit, thus additional gravitational assists were required. More information on the launch vehicle selection criteria can be found in Section~\ref{sec:LV}. By utilizing a \(\Delta V\)EGA leveraging orbit, similar to that of the Juno Mission~\cite{ret19}, Voyager 3’s launch C3 was reduced from 120 \(\text{km}^2 / \text{s}^2\) to 45 \(\text{km}^2 / \text{s}^2\), raising the mass limit to 17,600 kg. In this case, Voyager 3’s electric propulsion would provide the \(\Delta V\) for the deep-space maneuver (DSM) at aphelion. The advantage of this approach is that it only relies on Earth gravity assist(s), and so targeting specific flyby parameters is simplified. A 3:1\(^-\) Earth Leveraging orbit, discussed by Sims and Longuski in~\cite{ret20} was selected for this trajectory as it was a reasonable compromise between the amount of time spent interplanetary and required launch energy. It is important to note that the analysis covered by Sims and Longuski is designed to rendezvous with Jupiter. Voyager 3’s trajectory is aiming to maximize the incoming relative velocity.

After completing the flyby of Jupiter, the spacecraft is left with an offset from the solar focal line of 5.203 AU that must be made up for. On the path to 550 AU, Voyager 3 must close this distance and align itself with the solar focal line. To do so, 500 \( m/s \) of \( \Delta V \) must be held in reserve to reduce the velocity normal to the solar focal line to zero. As such, this reduces the outbound velocity of the spacecraft and adds an additional 4.5 months to the previously projected timeline. By budgeting for this fuel expenditure, it allows the spacecraft more time to collect the data along the solar focal line.

\subsubsection{Launch Vehicle Selection} \label{sec:LV}
{The trajectory design and weight of the spacecraft significantly limited the available launch vehicles. Because of the time of flight requirement, the trajectory requires a high launch C3 of 44.89 \(\text{km}^2 / \text{s}^2\) (\(V_\infty\) of 6.70 km/s). It is assumed that the launch vehicle would deliver this entirely in order to save the spacecraft’s propellant and to not have to use a kick stage. Existing Evolved Expendable Launch Vehicles (EELV) like ULA’s Delta IV Heavy and Atlas V 551 were considered for their reliability and payload-carrying capability. SpaceX's Falcon Heavy in the recoverable and fully expendable configurations were also investigated due to their cost. Still, this vehicle has only launched twice at the time of analysis. Finally, the NASA/Boeing Space Launch System (SLS) Block 1B and 2 were traded as these are heavy-lift, high-reliability vehicles designed for deep space missions with high launch energy requirements. Note that options such as Blue Origin’s New Glenn and SpaceX's Starship were not considered due to a lack of available data for a high energy launch. {ULA’s Vulcan was considered, but is not able to launch enough our design at a C3 of 44.89 \(\text{km}^2 / \text{s}^2\).} Starship and New Glenn could be prime contenders for this mission as they are intended to compete with the SLS for deep-space heavy-lift missions. { At the time of this analysis, both Starship and New Glenn do not have mass-C3 curves available for higher energy departures such as Voyager 3.}

The SLS is expected to launch NASA’s flagships in the coming future. Despite the launch cost and development status, the SLS Block 1B was chosen as it is the only physical vehicle able to meet the mass and energy requirements. Voyager 3 can be comfortably accommodated in the B8.4m-27000 fairing due to its large 7.5-meter diameter dynamic envelope and over 18-meter tall fairing~\cite{ret14}. Also, the spacecraft and launch vehicle adapter estimated to be <14,500 kg, {allowing for a 3,100 kg mass margin, giving ample room for mass growth over the course of the design process.} The SLS Block 1B could launch Voyager 3. 

\subsection{Spacecraft Design}
The Voyager 3 spacecraft itself is broken into three main sections, the bus which houses all of the electronics, communication equipment, and payload, the power assembly where the reactor and power distribution unit are, and finally the propulsion housing which contains the thrusters, controllers, and propellant. For a fully dimensioned view of the design, reference Figure~\ref{fig:retLayout} in the appendix.

\begin{figure}[htb]
  \centering
  \includegraphics[width = 0.666\textwidth]{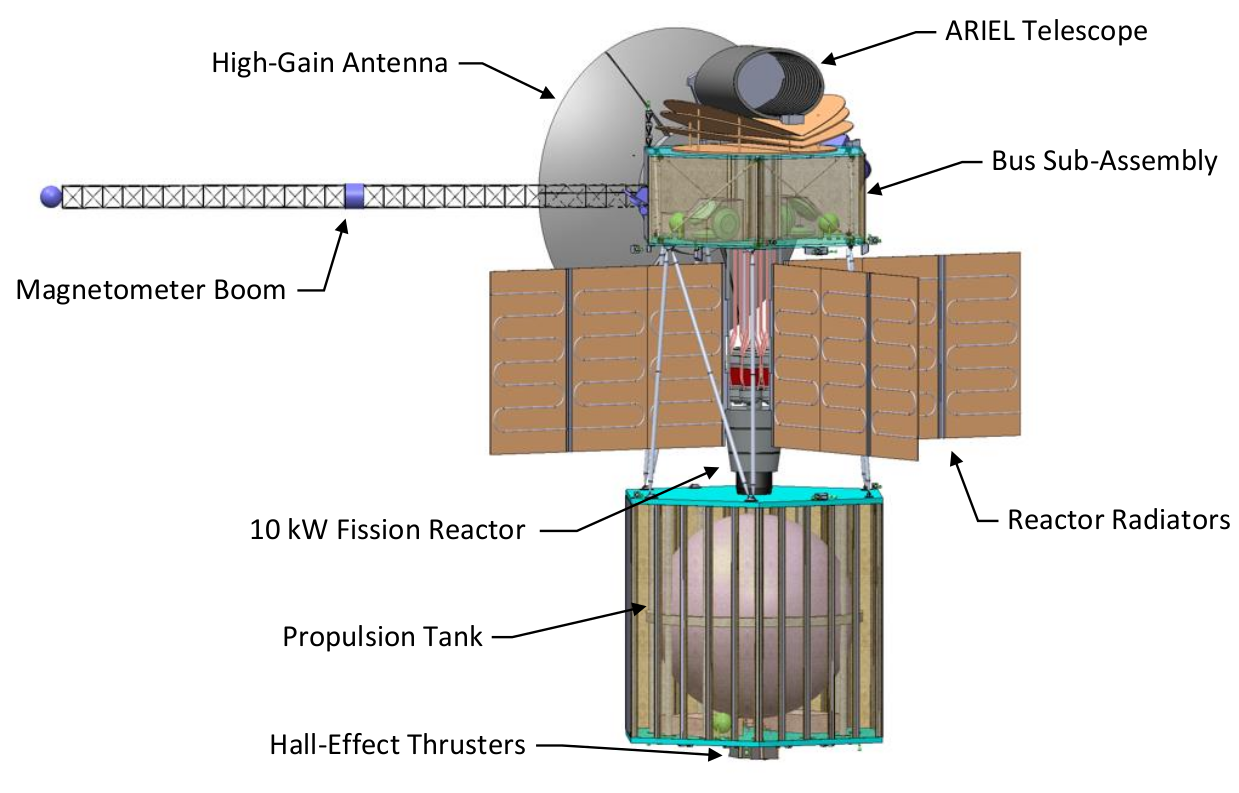}
  \caption{Voyager 3 component overview}
  \label{fig:retCompOverview}
\end{figure}

The origin point of the spacecraft's design lies with its sizable power system, the Kilopower Stirling nuclear reactor. While the use of radioisotope thermoelectric generators were considered, by taking into account the system power required, and the power generated per unit item, the RTG-focused power system was found to be incompatible with the requirements of the mission. At this point in time, nuclear reactor-powered spacecraft are relatively untested, as the only nuclear reactor ever flown by the United States was SNAP-10A in 1965~\cite{ret32}, there have been advances in the research of space-based nuclear reactors. NASA and the National Nuclear Security Administration, however, have been leading the design and testing of a new reactor technology called the Kilopower Reactor Using Sterling Technology (KRUSTY). This system is to undergo ground testing in the upcoming years toward a potential flight test on the lunar surface in the late 2020s.~\cite{ret33}. In this particular specification, the Kilopower reactor can output continuously 10kW of power. 

At the bottom of the spacecraft lies the propulsion modules. Based upon the requirements placed on the sub-system from the trajectory design, the engine must be able to provide 400mN of thrust continuously at an Isp greater than 2300 seconds. In addition to the performance requirements, the engine must be able to meet that max power limit of 8 kW. Engines with these performance specifications are currently in development, namely the Busek BHT-8000~\cite{ret30}, and will most likely have improved by the time Voyager 3 launches, allowing for additional performance gained from the trajectory. At these specifications, and the trajectories requirement of 32.4 km/s of \(\Delta V\), a total of 10,800 kg of xenon propellant would be required to deliver the spacecraft to 550 AU within 60 years.

One of the large challenges with traveling to 550 AU is being able to transfer large amounts of science data. The distance increases the size and power of the communication system required. Voyager 3 would use a 3.5 m high gain antenna for communication at high bit rates. The HGA is based on the DAWN spacecraft design albeit at a much larger size~\cite{ret62}. The spacecraft would also use three low gain antennas for the communication of data at lower bit rates to or from the spacecraft. The spacecraft utilizes the 34 m DSN antenna for both sending commands to the spacecraft and transmitting engineering and science data. Using the 34 m DSN antenna would make scheduling easier for the operations team. Voyager 3 would use X-Band communication and the DSN 34 m antenna is currently able to provide 68 {dBi \footnote{{dBi indicates the gain associated with a particular antenna receiver. It is based on the geometry and size of the antenna.}}} of gain in the X-Band range~\cite{ret60}. At a symbol rate of 58.4 ksps and transmitter power of 100 W, the datalink margin is 0.66 dB.

\begin{figure}[htb]
    \centering
    \includegraphics[width = 0.6\textwidth]{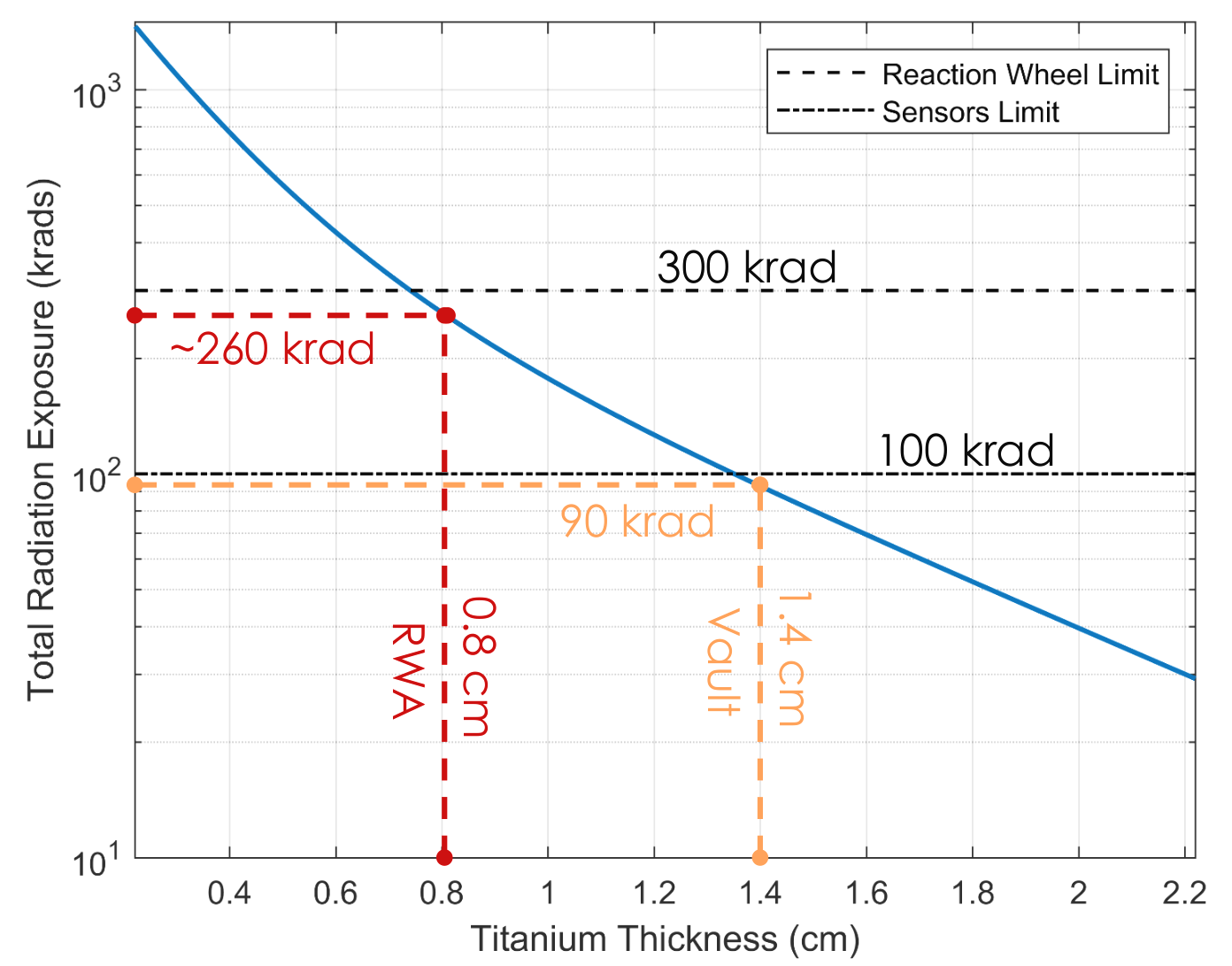}
    \caption{Total radiation dose from Jupiter flyby versus thickness of titanium radiation vault}
    \label{fig:retRadThickness}
\end{figure}

As the spacecraft would encounter many of the solar system’s harshest environments, including deep space and a low altitude flyby of Jupiter, the management of temperature and radiation would be paramount. Like Juno, within the bus would be a hardened radiation vault that would contain all sensitive electronics. The required areal density, parameterized by material thickness, can be seen demonstrated in Figure~\ref{fig:retRadThickness}. To keep warm in the coldest possible environment, the spacecraft would also utilize the heat from the nuclear reactor to protect its cold-sensitive components. Via conduction from heat pipes, the energy can be transmitted to warm up components in the extreme-cold environments. Likewise, the heat can be siphoned away from sensitive equipment towards the reactor radiator panels in the extreme-hot environments.

Attitude control system design is one of the major criteria for mission success as a telescoping mission. Historically, missions that required high levels of pointing accuracy have been plagued by issues of reaction wheel failure in the past~\cite{CasAttitude}. {Voyager 3 would encounter these issues as well since reaction wheels are necessary for high pointing accuracy missions.} For the spacecraft to withstand the long mission time the spacecraft must have a way of mitigating the damage done to the reaction wheel. The primary method of failure mitigation for the attitude control system is redundancy. This results in the spacecraft having 4 primary reaction wheel assemblies (RWAs) with 4 backup RWAs. The reason for this is two-fold, reaction wheels are notorious for have short lifespans and reaction wheels have failed before the main portion of the spacecraft mission in the past~\cite{CasAttitude}. The second strategy is limiting the use of the reaction wheels after Jupiter to periodic equipment checkouts and science objectives. By limiting the use of the reaction wheels the number of revolutions the reaction wheels would experience {are minimized. Furthermore, attitude control thrusters are included in the design; however, they would not provide enough accuracy to utilize the telescope.}   

With all of the major subsystems defined, a total of launch mass and power requirements can be tallied in Table~\ref{tab:retMassPower}. Values of note for the mass include the power system allocation of 1445 kg, primarily for the 10kW Kilopower nuclear reactor and its accompanying cooling systems. Additionally, for the spacecraft power usage, the propulsion section stands out for the 8kW of power required for the Hall-Effect thruster operation. Broken down in Table~\ref{tab:retCost}, is the associated estimated cost of the Voyager 3 program. Of note is the operations cost due to the extreme mission length and continual required downlinks. {Cost estimation was performed by NASA's PCEC, and primarily used as an estimate to be further refined over the course of the design. The high cost of the operations is one such limitation of the software, which is not well equipped to handle such long mission durations.}

\begin{table}[htb]
  \centering
  \begin{minipage}{0.45\textwidth}
  \caption{Total mass and power of primary Voyager 3 architecture}
  \label{tab:retMassPower}
  \begin{tabular}{c c c}
    \hline
    \textbf{Sub System}          & \textbf{Mass (kg)} & \textbf{Power (W)} \\ \hline
    Structures         & 440                & 10          \\ \hline
    Radiation         & 240                & 0           \\ \hline
    Thermal           & 50                 & 10          \\ \hline
    ADACS             & 150                & 125         \\ \hline
    Power             & 1,445\footnotemark[1] & 120         \\ \hline
    Cabling                      & 80                 & 0           \\ \hline
    Propulsion                   & 290                & 8,000        \\ \hline
    Telecomm.           & 50                 & 395         \\ \hline
    C\&DH     & 75\footnotemark[1] & 55          \\ \hline
    Sub-Total & 3,000\footnotemark[2] & 9,440\footnotemark[2] \\ \hline
     \quad Margins & 15\%\footnotemark[3] & 20\% \\ \hline
    Payload                      & 200                & 150         \\ \hline
    Propellant                   & 10800              & 0           \\ \hline
     Totals   & 14,000    & 9,590 W  \\ \hline
  \end{tabular} 
  \end{minipage}%
  \hfill
  \centering
  \begin{minipage}{0.45\textwidth}
  \caption{Total flight cost of Voyager 3, including development, construction, and mission operations.}
  \label{tab:retCost}
  \begin{tabular}{cc}
    \hline
    \textbf{Component}                     & \textbf{\$M (FY2020)} \\ \hline
    Project Management                     & 18.51                 \\ \hline
    Systems Engineering                    & 65.47                 \\ \hline
    Mission Assurance                      & 18.43                 \\ \hline
    Science/Technology                     & 17.32                 \\ \hline
    Payload                                & 26.4                  \\ \hline
    Flight System                          & 671.28                \\ \hline
    Mission Operations                     & 1,248.72              \\ \hline
    Launch Services w/o LV                 & 195.08                \\ \hline
    Ground Data Systems                    & 168.24                \\ \hline
    System Integration Test,\\ \& Checkout & 21.08                 \\ \hline
    Outreach                               & 0.56                  \\ \hline
 Total           & 2,451.07      \\ \hline
  \end{tabular}
  \end{minipage}%
\end{table}

\footnotetext[1]{Masses for the subsystems are those that were not considered for the additional margin. These systems are considered off-the-shelf components, thus their mass and dependencies do not change}
\footnotetext[2]{Below margins included in sub-total values}
\footnotetext[3]{The mass margin value of 15\% is the recommended number by AIAA for spacecraft over 2500 kg and at a PDR stage of design~\cite{ANSIAIAA}.}

\clearpage

\section{Conclusion}
The Voyager 3 mission is only proposed with current or in development technologies. By the time of the proposed mission launch, technology will likely be even further developed. The Voyager 3 mission concept proposes to perform direct imaging of an exoplanet such as TRAPPIST-1 star system using the solar gravitational lensing. The science requirements and the mission concept were discussed in this paper, and the instrument suite was reviewed.
While the proposed 60-year duration of the mission is beyond standard mission practices, it is in itself a significant demonstration. The Voyager 3 mission concept study has demonstrated that a mission to 550 AU is not only feasible, within a reasonable time, but also can return significant scientific information, providing compelling science on the way to, at, and from the ISM, and demonstrating enabling technologies towards reaching another star. 

The spacecraft would be equipped with a wide-angle camera, {magnetometer}, {solar wind spectrometer}, {dust analyzer}, {high energy particle spectrometer}, and a telescope with a main mirror diameter at or larger than 1 meter.  
This instrument suite allows Voyager 3 to measure the Extragalactic Background Light and Zodiacal light pollution, and study the Interplanetary Dust \& Plasma. Voyager 3 could also extend the baseline of parallax measurements to 550 AU allowing for the most accurate measurements of the Hubble constant $H_0$ to date, and make observations of small bodies in our outer solar system.  
To accomplish these goals, Voyager 3's primary architecture's use of nuclear electric propulsion allows the spacecraft to reach speeds in excess of 3 times the current fastest spacecraft leaving the solar system, Voyager 1, at 11.1 AU/yr. In conjunction with its suite of scientific equipment to study the solar system and beyond, the proposed system has the capability to reach the gravitational lensing boundary of 550 AU in 58 years, and be able to continue on the solar focal line, enabling the extended collection of exoplanetary data. In addition, the spacecraft's trajectory allows the scientific community to gain access to the payload equipment during its 38-year cruise phase. 

{Additional studies were conducted on the viability of low-perihelion Oberth maneuvers which yielded promising results, and new science opportunities not available to the current chosen design architecture. As discussed in Section 1, the required technologies were too far away from being a possible mission option for this time frame.}



\newpage

\section*{Appendix}

\subsection{Mapping Science Goals}
The majority of the science goals and objectives for the Voyager 3 mission are selected based on the NASA Strategic Objectives document as well as Planetary Science Decadal Survey~\cite{P1}. Voyager 3 responds to NASA 2018 Strategic Plan, by including two major objectives:
Objective 1.1: “Understand the Sun, Earth, Solar System, and Universe,” which calls for proposals related to planetary science and understanding the making of the universe, conducting scientific studies of the Sun from space and returning data from other bodies in the solar system. Objective 3.1: “Develop and Transfer Revolutionary Technologies to Enable Exploration Capabilities for NASA and the Nation,” which calls for increasing access to planetary surfaces. 

Voyager 3 investigates science objective identified by the Planetary Science Decadal Survey under the following questions:
\begin{itemize}
\item Priority Question 1: What were the initial stages, conditions, and processes of solar system formation and the nature of the interstellar matter that was incorporated? 
\item Priority Question 2: How did the giant planets and their satellite systems accrete, and is there evidence that they migrated to new orbital positions? 
\item Priority Question 4: The planetary habitats: What were the primordial sources of organic matter, and where does organic synthesis continue today? 
\item Priority Question 6: Beyond Earth, are there modern habitats elsewhere in the solar system with necessary conditions, organic matter, water, energy, and nutrients to sustain life, and do organisms live there now? 
\item Priority Question 7: How do the giant planets serve as laboratories to understand Earth, the solar system, and extrasolar planetary systems? 
\end{itemize} 
Voyager 3 also expands our knowledge by addressing key questions in the Astronomy and Astrophysics Decadal Survey:
\begin{itemize}
\item Cosmology and Fundamental Physics (CFP) 1: How did the universe begin?	
Cosmology and Fundamental Physics (CFP) 2: Why is the universe accelerating?
\item Cosmology and Fundamental Physics (CFP) 3: What is dark matter?
\item Galaxies Across Cosmic Time (GCT) 4: What were the first objects to light up the universe, and when did they do it?
\item Planetary Systems and Star Formation (PSF) 2: How do circumstellar disks evolve and form planetary systems?
\item Planetary Systems and Star Formation (PSF) 4: Do habitable worlds exist around other stars, and can we identify the telltale signs of life on an exoplanet?
\end{itemize} 
These objectives and questions are mapped to our mission science objectives in Table~\ref{tab:mapping science}. 

\begin{table}[htb]
\caption{Mapping Science Goals}
  \label{tab:mapping science}
   \centering
  \footnotesize
 \begin{tabular}{ p{0.35\textwidth} p{0.1\textwidth} p{0.2\textwidth} p{0.2\textwidth}  }
 \hline
 \multicolumn{1}{c}{Voyager 3 Science Objective}&{NASA 2018 Strategic Plan }&{Vision and Voyages for Planetary Science in the Decade 2013-2022}&{The Astronomy and Astrophysics Decadal Survey}\\
 \hline
SO1.1: Jupiter Flyby& Objective 3.1 & Priority Question 2, 7 &    \\  \hline
SO1.2: Solar Magnetic Fields &Objective 1.1 && \\ \hline
SO1.3: Solar Winds & Objective 1.1 &&\\ \hline
SO1.4: Zodiacal Dust cloud &Objective 1.1 &&PSF Question 2\\ \hline
SO1.5: Extragalactic Background Light (EBL) &Objective 1.1 &&GCT Question 4\\ \hline
SO1.6: Small Body Science & Objective 3.1 & Priority Questions 1, 4, 6 &\\ \hline
SO2.1: Interplanetary Dust and Plasma &Objective 1.1 &Priority Question 1  &PSF Question 2\\ \hline
SO2.2: Parallax Science &Objective 1.1&&CFP Questions 1, 2\\ \hline
SO3.1: Solar Gravitational Lensing (SGL) &Objective 1.1 &Priority Question 6 &PSF Question 4\\
 \hline
\end{tabular}
  \end{table}
\subsection{Science Traceability Matrix}
A science traceability matrix (STM) is used to ensure Voyager 3's
instrument suite and mission plan will be able to meet the
science goals presented in Table~\ref{tab:scienceTraceMatrix}.

\begin{table}[htb]
  \footnotesize
  \caption{Science Traceability Matrix}
  \label{tab:scienceTraceMatrix}
  \centering
  \begin{tabular}{p{1in} p{2in} p{3in}}
  \hline
  \multicolumn{2}{c}{Science Objective}&{Measurements Objectives}\\ 
  \hline
  SO1.1: Jupiter Flyby&
   \begin{itemize} \item  Investigate the chemistry of giant planet atmospheres 
  \item Search for possible evidence of planetary migration \item Evaluate solar wind and magnetic field interaction with Jupiter \end{itemize}&
   \begin{itemize} \item Measuring non-thermal radio emissions and energetic atoms 
  \item Direct detection of orbital expansions for the inner Jovian moon 
  \item Determining the characteristics of Jovian impacts 
  \item Measuring key elemental and isotropic abundances and thermal profiles in the atmospheres of Jupiter \item Mapping the magnetic field of Jupiter during flyby \end{itemize}  \\ \hline
  SO1.2: Solar Magnetic Fields \&  SO1.3: Solar Winds &
   \begin{itemize} \item Characterize and determine boundaries of termination shock, heliosheath, and heliopause
  \item Measure superthermal pick-up ion (PUI) population, as well as energetic electrons \end{itemize}&\begin{itemize} \item Track solar wind evolution from 5 solar radii out to edge of HP
  \item Search for boundaries between regions of TS, HS, and HP
  \item Measuring PUI population and energetic electrons in the range of 1-40 keV\end{itemize}
  \\ \hline
  SO1.4: Zodiacal Dust Cloud&
  \begin{itemize} \item Study zodiacal light intensity as function of heliocentric distance \item Model diffusion rate of dust from planetesimals throughout solar system \item Map the dust density and composition as a function of radial position \item Understand the dust population and compare to dust clouds of other extrasolar systems\end{itemize}&\begin{itemize} \item Measuring zodiacal light intensity as a function of heliocentric distance to at least 3 AU with uncertainty $<5\%$
  \item Measuring IPDC grain size, density, albedo as a function of distance to at least 100 AU with uncertainty $<1\%$\end{itemize}\\ \hline
  SO1.5: Extragalactic Background Light (EBL) &\begin{itemize} \item Obtain Extragalactic Background Light (EBL) spectrum and brightness for optical and infrared (IR) wavelengths 
  \item Observe EBL beyond 10 AU to reduce effect of zodiacal light pollution \end{itemize}&\begin{itemize} \item Measuring EBL intensities $<0.8$ nW/(m$^2$Sr) with uncertainty $< 1\%$
  \item Minimum heliocentric distance of 10AU to eliminate zodiacal light effect
  Requires measurement range of $\lambda=1-18 \mu m$ \end{itemize} \\  \hline
  SO1.6: Small Body Science &\begin{itemize} \item Image objects of interest in the Main Asteroid Belt and the Kuiper Belt in high resolution
  \item Detect, track, and characterize both Kuiper Belt Objects and Main Asteroid Belt Objects \end{itemize}& \begin{itemize} \item Capturing images of asteroids/KBOs with a minimum resolution of 200 m/ pixel \item Using visible imaging for hot objects 
  \item Using IR imaging for cold objects\end{itemize}  \\
  \hline
  SO2.1: Interplanetary Dust and Plasma &\begin{itemize} \item Determine dust composition 
  \item Understand dust kinematics
  \item Study grain size distribution throughout ISM
  \item Model the plasma density variations in ISM\end{itemize}&\begin{itemize} \item Measure dust composition
  \item Measure grain size distribution
  \item Measure dust impact rates
  \item Measure plasma temperatures
  \item Measure plasma density\end{itemize} \\ \hline
  SO2.2: Parallax Science & \begin{itemize} \item Taking advantage of a baseline up to 550 AU
  \item Make the most accurate measurements of the Hubble constant $H_0$ to date\end{itemize}& \begin{itemize} \item Requires a second simultaneous observation from Earth to eliminate the effect of stellar motion\item Measure objects with distances $> 10$ Mpc
  \item Constrain $H_0$ with uncertainty $<1\%$\end{itemize}\\ 
  \hline
  SO3.1: Solar Gravitational Lensing (SGL) & \begin{itemize} \item Obtain direct images of the exoplanets around the star Wolf 1061 by utilizing the gravitational lensing of the Sun
  \item  Search for biosignatures in exoplanet atmospheres \end{itemize} & \begin{itemize} \item Requires spacecraft minimum heliocentric distance of 550 AU \item Requires Pointing precision must be maintained within $\sim1-10 \mu as$ \item Obtain spectroscopic images over a minimum wavelength range of $0.1-20 \mu m$
  \item Search for atmospheric biosignatures: $O_2$, $O_3$, $N_2O$, $NH_3$, $CH_4$, $C_2H_6$\end{itemize}\\
  \hline
  \end{tabular}
  \end{table}

\subsection{Measurement Objectives}
Voyager 3 will investigate each science objective by a series of measurements listed in Table~\ref{tab:measurementObjectives}. The table also provides the science instruments needed for proposed measurements on the spacecraft. Our approach includes combining a set of 7 instruments on the spacecraft.  
  
\begin{table}[htb]
\footnotesize
\caption{Measurement Objective and Instruments}
\label{tab:measurementObjectives}
\centering
\begin{tabular}{ p{.15\textwidth} p{.25\textwidth} p{.1\textwidth} p{.1\textwidth}p{.25\textwidth} }
\hline
~&{Measurements Objectives}&{Measurement Instrument}&{Instrument Performance} &{Mission Functional Requirement}\\\hline
\multirow{1}{.1\textwidth}{SO1.1: Jupiter Flyby}  & Atmosphere's composition; Elemental and isotropic abundance& 
    CDA& $10^{-15}-10^{-6} $kg& Trajectory through Jupiter's magnetosphere \\
\cline{2-5}
  & Distribution of asteroids& ARIEL; Wide angle camera &$<0.1\mu m$& \\
\cline{2-5}
& Imaging of the moons & ARIEL; Wide angle camera &$<0.1\mu m$& \\
\cline{2-5}
   & Magnetic field & MAG-3 & $100 \mu V/nT$& Trajectory
through
Jupiter's
magnetosphere\\
   \cline{2-5}
   & Solar wind; Energetic atoms  & SWICS &$100 \mu V/nT$& Trajectory
through
Jupiter's
magnetosphere\\
\hline 
SO1.2: Solar Magnetic Fields & Magnetic field's map; Magnetic field’s influence on interstellar matter & MAG-3& 
    $100 \mu V/nT$&Distance: 1AU-- edge of HP\\
 \hline
SO1.3: Solar Winds& evolution of HP; boundaries of TS, HS, and HP; PUI population& SWICS&  $0.16 – 70 keV$ & Distance: 1AU-- edge of HP\\
\hline
\multirow{1}{.1\textwidth}{SO1.4: Zodiacal Dust Cloud} & Zodiacal light intensity Vs heliocentric distance& ARIEL&  $<0.1\mu m$; $\lambda=0.03-18 \mu m$& Distance: 10--$>$100AU\\
\cline{2-5}
  & Diffusion rate of dust; Dust density map and composition; Dust population & CDA &$10^{-15}-10^{-6} $kg&Distance: 10--$>$100AU \\
\hline
SO1.5: EBL& EBL intensities & ARIEL& $\lambda$=0.03-18$\mu$m & Distance$>$
10AU \\
\cline{2-5}
\hline
\multirow{1}{.1\textwidth}{SO1.6: Small Body Science}  & Images of main asteroid belt objects  & ARIEL& $\lambda$=0.03-18$\mu$m & Distance$>$2AU \\
\cline{2-5}
  & Images of KBOs& ARIEL& $\lambda$=0.03-18$\mu$m& Distance: 30-50 AU \\
\cline{2-5}
\hline 
\multirow{1}{.1\textwidth}{SO2.1: Interplanetary and ISM Dust}  & Dust composition& CDA& $10^{-15}-10^{-6}$kg & Distance$>$2AU \\
\cline{2-5}
  & Dust kinematics& CDA& $10^{-15}-10^{-6}$kg& Distance: 30-50 AU \\
\cline{2-5}
  & Grain size distribution& CDA& $10^{-15}-10^{-6}$kg& Distance: 30-50 AU \\
\cline{2-5}
\hline 
\multirow{1}{.1\textwidth}{SO2.2: Parallax Science} & Angular position of stars & ARIEL& Angular resolution: 7.5$\times 10^{-3}$ as 
   &Distance$>$40AU, Requires a second simultaneous observation from Earth to eliminate the effect of stellar motion\\
\cline{2-5}
 & Hubble constant $H_0$ & ARIEL & Angular resolution: 7.5$\times 10^{-3}$ as & \\
\cline{2-5}
\hline 
\multirow{1}{.1\textwidth}{SO3.1: SGL} & Direct image of the exoplanets & ARIEL& $\lambda$=0.03-18$\mu$m
     &Distance$>$ 547 AU\\
\cline{2-5}
 & Biosignatures in exoplanet atmospheres & ARIEL & $\lambda$=0.03-18$\mu$m &Distance$>$ 547 AU\\
\cline{2-5}
\hline
\end{tabular}
\end{table}

\clearpage

\subsection{3-View Dimensioned Drawings}

\begin{center}\rotatebox{90}{\begin{minipage}{0.95\textheight}
    \centering
    \includegraphics[width = \textwidth]{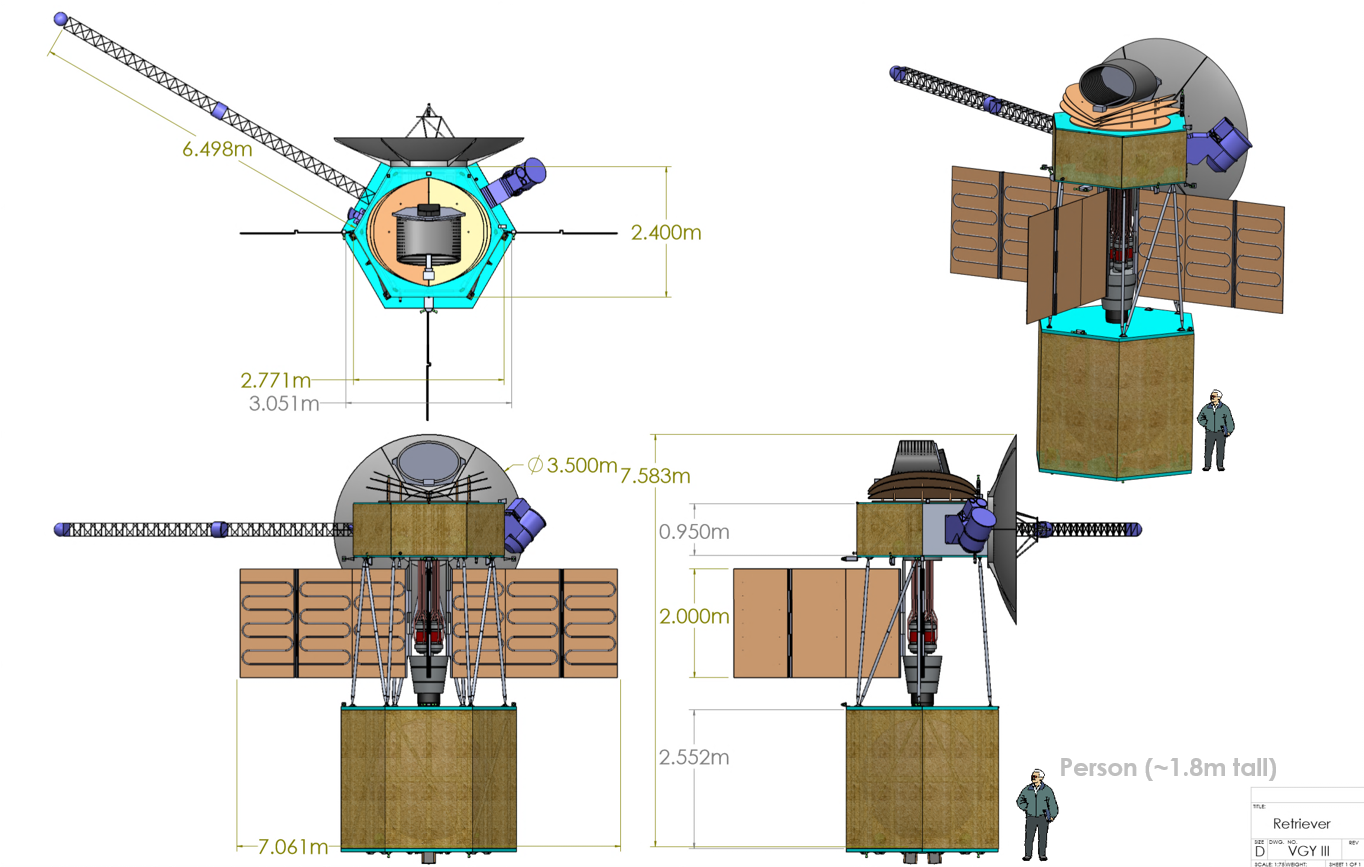}
    \captionof{figure}{Dimensioned 3-view of entire spacecraft architecture (1.8m human for scale)}
    \label{fig:retLayout}
  \end{minipage}}\end{center}



\section*{Acknowledgments}

This research was conducted over 2 years as an undergraduate research project at Cal Poly Pomona. The team was advised by Dr. Navid Nakhjiri, Dr. Shohreh Abdolrahimi, and Dr. Christos Tzounis at Cal Poly Pomona and Dr. David Scott at NASA JPL. We thank Dr. Ed C. Stone, James A. Smith, Jacob W. Smith, and Emilio L. Vazquez who proposed the initial concept for this mission. We also thank Dr. Try Lam and Dr. Damon Landau who helped the students with the trajectory design of this project. Additionally, we thank spacecraft design reviewers Stephen Edberg, Dr. Don Edberg and Steve Nanning for their constructive feedback on various parts of the design. 

Additional Students who contributed to this project include: Ricardo Cancinos, Cory B. Hassel, Angel Linares, Demetria Ma, Garret Parham, Ricardo Paz, Jessie Portillo, Aayush Sareen, Noah Tonies, Patrick Behr, Phillip Campos, Leara M. Domingueza, and Tim M. Kilduff

 The research was carried out at the Jet Propulsion Laboratory, California Institute of Technology, under a contract with the National Aeronautics and Space Administration (80NM0018D0004). The information presented about the Voyager 3 mission concept is pre-decisional and is provided for planning and discussion purposes only. The cost information contained in this document is of a budgetary and planning nature and is intended for informational purposes only. It does not constitute a commitment on the part of JPL and/or Caltech.



\begin{thebibliography}{}

\bibitem{P22} E. Stone, L. Alkalai,  and L. Friedman, "Science and enabling technologies to explore the interstellar medium," Keck Institute for Space Studies (2015), DOI: 10.26206/3PKN-N663.  
\bibitem{P1} Committee on the Planetary Science Decadal Survey; National Research Council, "Vision and Voyages for Planetary Science in the Decade 2013-2022," National Academic Press, Washington, D. C. (2013), DOI: https://doi.org/10.17226/13117. 
\bibitem{P23} M. Opher, E. C. Stone, and P. C. Liewer, "The effects of a local interstellar magnetic field on Voyager 1 and 2 observations," The Astrophysical Journal Letters, 640:L71 (2006), DOI: https://doi.org/10.1086/503251. 
\bibitem{P24} A. Michael, M. Opher, E. Provornikova, J. Richardson, and G. Toth, "Magnetic Flux Conservation in the Heliosheath Including Solar Cycle Variations of Magnetic Field Intensity," The Astrophysical Journal Letters, 803:L6 (2015), DOI: https://doi.org/10.1088/2041-8205/803/1/L6.
\bibitem{P25} G. Gloeckler, L. A. Fisk, and L. J. Lanzerotti, "Acceleration of solar wind and pickup ions by shocks," Proceedings in Solar Wind 11/SOHO 16, Connecting Sun and Heliosphere, Conference (2005), {\bf bibcode: 2005ESASP.592..107G}. 
\bibitem{P26} Kelsall, T., et al. "The COBE diffuse infrared background experiment search for the cosmic infrared background. II. Model of the interplanetary dust cloud." The Astrophysical Journal 508.1 (1998): 44, {\bf bibcode: 2020arXiv200508940P}.
\bibitem{P27} G. Stenborg, J. R. Stauffer, and R. A. Howard, "Evidence for a Circumsolar Dust Ring Near Mercury's Orbit," The Astrophysical Journal, 868:74 (2018), DOI: https://doi.org/10.3847/1538-4357/aae6cb.
\bibitem{P28} C. M. Casey, D. Narayanan, and A. Cooray, "Dusty star-forming galaxies at high redshift," Physics Reports, 541:45 (2014), DOI: https://doi.org/10.1016/j.physrep.2014.02.009. 
\bibitem{P29} A. Cooray, "Extragalactic Background Light Measurements and Applications," Royal Society Open Science 3:3 (2016), DOI: https://doi.org/10.1098/rsos.150555.
\bibitem{ZEBRA} https://www.universetoday.com/tag/zebra-zodiacal-dust/. [Accessed 2021].
\bibitem{Hauser2001} Hauser, Michael G., and Eli Dwek. "The cosmic infrared background: measurements and implications." Annual Review of Astronomy and Astrophysics 39.1 (2001): 249-307, DOI: https://doi.org/10.1146/annurev.astro.39.1.249.
\bibitem{Matsuoka} Matsuoka, Yoshiki, et al. "Cosmic optical background: The view from Pioneer 10/11." The Astrophysical Journal 736.2 (2011): 119, DOI: https://doi.org/10.1088/0004-637X/736/2/119.
\bibitem{Levenson2007} Levenson, L. R., E. L. Wright, and B. D. Johnson. "DIRBE minus 2MASS: confirming the CIRB in 40 new regions at 2.2 and 3.5 $\mu$m." The Astrophysical Journal 666.1 (2007): 34, DOI: https://doi.org/10.1086/520112.
\bibitem{Hauser} Hauser, Michael G., et al. "The COBE diffuse infrared background experiment search for the cosmic infrared background. I. Limits and detections." The Astrophysical Journal 508.1 (1998): 25, DOI: https://doi.org/10.1086/306379.
\bibitem{Kelsall:1998bq} Kelsall, T., et al. "The COBE diffuse infrared background experiment search for the cosmic infrared background. II. Model of the interplanetary dust cloud." The Astrophysical Journal 508.1 (1998): 44.
\bibitem{P15} A. C. Barr and R. M. Canup, "Origin of the Ganymede-Callisto dichotomy by impacts during the late heavy bombardment," Nature Geoscience 3:164-167 (2010), DOI: https://doi.org/10.1038/ngeo746.
\bibitem{P16} Jacobsen, Benjamin, et al. "26Al–26Mg and 207Pb–206Pb systematics of Allende CAIs: canonical solar initial 26Al/27Al ratio reinstated." Earth and Planetary Science Letters 272.1-2 (2008): 353-364, DOI: https://doi.org/10.1016/j.epsl.2008.05.003.
\bibitem{P18} Brownlee, Don, et al. "Comet 81P/Wild 2 under a microscope." science 314.5806 (2006): 1711-1716, DOI: 10.1126/science.1135840.
\bibitem{P19} Bernstein, Gary M., et al. "The size distribution of trans-Neptunian bodies." The Astronomical Journal 128.3 (2004): 1364 DOI: https://doi.org/10.1086/422919.
\bibitem{P20} K. Tsiganis, R. Gomes, A. Morbidelli, and H.F. Levison, "Origin of the orbital architecture of the giant planets of the solar system," Nature 435:459-461 (2006), DOI: https://doi.org/10.1038/nature03539. 
\bibitem{P21} National Research Council, "New Worlds, New Horizons in Astronomy and Astrophysics," The National Academies Press, Washington, D.C. (2010), DOI: https://doi.org/10.17226/12982.
\bibitem{P33} M. G. Wolfire, C. F. McKee, D. Hollenbach, A. and Tielens, "Neutral atomic phases of the interstellar medium in the galaxy," The Astrophysical Journal, 587:278 (2003), DOI: https://doi.org/10.1086/368016. 
\bibitem{P34} D. J. Schlegel, D. P. Finkbeiner, M. Davis, "Maps of dust infrared emission for use in estimation of reddening and cosmic microwave background radiation foregrounds," The Astrophysical Journal 500: 525 (1998) DOI: https://doi.org/10.1086/305772.
\bibitem{P36} A. Jenkins, R. Villard, and A. Riess, "NASA: Hubble and Gaia Team Up to Fuel Cosmic Conundrum," NASA, https://www.nasa.gov/feature/
\bibitem{P37} A. Riess, "New Parallaxes of Galactic Cepheids from Spatially Scanning the Hubble Space Telescope: Implications for the Hubble Constant," The Astrophysical Journal 855: 18 (2018), DOI: https://doi.org/10.3847/1538-4357/aaadb7.
\bibitem{P39} Committee on a Decadal Strategy for Solar and Space Physics (Heliophysics), National Research Council, "Solar and Space Physics A Science for a Technological Society," The National Academies Press, Washington, D. C. (2009), DOI: 10.17226/13060.
\bibitem{Eddington} Dyson, Frank Watson, Arthur Stanley Eddington, and Charles Davidson. "IX. A determination of the deflection of light by the Sun's gravitational field, from observations made at the total eclipse of May 29, 1919." Philosophical Transactions of the Royal Society of London. Series A, Containing Papers of a Mathematical or Physical Character 220.571-581 (1920): 291-333, DOI: https://doi.org/10.1098/rsta.1920.0009.
\bibitem{Kraus}  J. D. Kraus, "Radio astronomy (2nd ed.)," Cygnus-Quasar Books, 1986.
\bibitem{Turyshev} S. G. Turyshev and B. Andersson, "The 550-AU mission: a critical discussion," Monthly Notices
of the Royal Astronomical Society, vol. 341, pp. 577-582, 2003. https://doi.org/10.1046/j.1365-8711.2003.06428.x
\bibitem{R7} Carroll, Ostlie. "The Interstellar Medium and Star Formation. An Introduction to Modern Astrophysics," 2nd Ed.  Pearson 2014. 
\bibitem{Landis} G. Landis, "A Telescope at the Solar Gravitational Lens," JBIS, Vol 71, 369 (2018). 
\bibitem{TURYSHEV2018} Turshev et. al., "Direct Multipixel Imaging of an Exo-earth with a Solar Gravitational Lens Telescope," JBIS, Vol 71, 361 (2018). 
\bibitem{TuryshevReport} S. G. Turyshev et al., "Directy Multiplixel Imaging and Spectroscopy of an Exoplanet with a Solar Gravity Lens Mission," NASA Innovative Advanced Concepts (NIAC) Phase I. 
\bibitem{Turyshev2020} Turyshev, Slava G., and Viktor T. Toth. "Image formation for extended sources with the solar gravitational lens." Physical Review D 102.2 (2020): 024038. DOI: 10.1103/PhysRevD.102.024038. 
\bibitem{ret1} NASA JPL, "Voyager Did You Know," [Online]. Available: https://voyager.jpl.nasa.gov/mission/did-you-know/. [Accessed April 2020].
\bibitem{ret2} J.R. C. Cook, "Juno Armored Up to Go to Jupiter," NASA, 12 August 2010. [Online]. Available: https://www.nasa.gov/mission\_pages/juno/news/juno20100712.html. [Accessed May 2020].
\bibitem{ret3} NASA Technology Transfer Program, "Project Cost Estimating Capability (PCEC)," NASA, May 2020. [Online]. Available: https://software.nasa.gov/featuredsoftware/pcec. [Accessed May 2020].
\bibitem{ret4} NASA, "The Planetary Society," [Online]. Available: https://www.planetary.org/blogs/casey-dreier/images/nasa-flight-mission-project-life-cycle.html.
\bibitem{ret5} R. A. T. a. C. H. Jaroschek, "The Heliospheric Termination Shock," 25 July 2008. [Online]. Available: \\ https://arxiv.org/pdf/0807.4170.pdf. [Accessed 2020].
\bibitem{ret7} G. A. Landis, "Mission to the Gravitational Focus of the Sun:," [Online]. Available: \\ https://arxiv.org/ftp/arxiv/papers/1604/1604.06351.pdf. [Accessed 2020].
\bibitem{ret9} SpaceQuest, Ltd., " MAG-3 Satellite Magnetometer," Fairfax.
\bibitem{ret10} G. Gloecker, "The Solar Wind Ion Composition Spectrometer," Astronomy and Astrophysics Supplement Series, vol. 92, no. 2, pp. 267-289, 1992. 1992A\&AS...92..267G.
\bibitem{ret11} S. R, A. N and L. D, "The Cassini Cosmic Dust Analyzer," Kluwer Academic, Netherlands, 2004. DOI: 10.1007/978-1-4020-2774-1-7. 
\bibitem{ret12} P. Eccleston and K. Middleton, "ARIEL Payload Design Description," RAL Space, 2017.
\bibitem{ret14} National Aeronautics and Space Administration, "Space Launch System (SLS) Program Vehicle Data Summary," [Online]. Available: https://ntrs.nasa.gov/archive/nasa/casi.ntrs.nasa.gov/20120014495.pdf.
\bibitem{ret15} N. Arora, N. Strange and L. Alkalai, "Trajectories for a Near Term Mission to the Interstellar Medium," JPL Technical Report Server, Pasadena, 2015.
\bibitem{ret17} D. Landau and T. Lam, "Broad Search and Optimization of Solar Electric Propulsion Trajectories to Uranus and Neptune,"Advances in Astronautical Sciences, vol. 135, pp. 2093-2112, 2010. DOI: 10.34133/2021/5245136. 
\bibitem{ret19} T. A. Pavlak, R. B. Frauenholz, J. J. Bordi, J. A. Kangas and C. E. Helfrich, "Maneuver Design for the Juno Mission: Inner Cruise," American Institute of Aeronautics and Astronautics, 2014.
\bibitem{ret20} J. A. Sims and J. M. Longuski, "Analysis of \( V_\infty \) Leveraging for Interplanetary Missions," American Institute of Aeronautics and Astronautics, West Lafayette, 1994.
\bibitem{ret30} I. Busek Co., "GEOSAT Propulsion System Architecture with Electric Apogee Motor". United States of America Patent 15/138,748, 22 December 2016.
\bibitem{ret32} Idaho National Laboratory, "Atomic Power in Space II," 2015.
\bibitem{ret33} M. A. Gibson, L. Mason and C. Bowman, "Development of NASA's Small Fission Power System for Science and Human Exploration," American Institute of Aeronautics and Astronautics, Oak Ridge.
\bibitem{ret34} M. A. Gibson, D. I. Poston and P. McClure, "NASA’s Kilopower Reactor Development and the Path to Higher Power Missions," NASA.
\bibitem{ret35} J. F. Zakrajsek, D. F. Woerner and J.-P. Fleurial, "NASA Special Session: Next-Generation Radioisotope Thermoelectric Generator (RTG) Discussion," NASA.
\bibitem{ret38} M. V. Podzolko, I. V. Getselev, Y. I. Gubar, I. S. Veselovsky and A. A. Sukhanov, "Charged Particle Fluxees and Radiation Does in Earth-Jupiter-Europa Spacecraft's Trajectory," February 2009. [Online]. Available: http://www.iki.rssi.ru/conf/2009elw/presentations/presentations|\_pdf/session2/podzolko|\_getselev\_ELW.pdf. [Accessed 2020].
\bibitem{ret39} Honeywell, "GG1320AN Digital Laser Gyro," [Online]. Available: https://aerospace.honeywell.com/content/dam/aero/en-us/documents/learn/products/sensors/brochures/GG1320ANDigitalLaserGyro-bro.pdf. [Accessed 2020].
\bibitem{ret40} Honeywell, "Constellation Series Reation Wheels," Honeywell.
\bibitem{ret60} JPL, "Deep Space Network Services Catalog," 2015.
\bibitem{ret62} J. Taylor, "Dawn Telecommunications: DESCANSO Design and Performance Summary Series," JPL.
\bibitem{ret63} J. Taylor, "Cassini Orbiter/Huygens Probe Telecommunications: DESCANSO Design and Performance Summary Series,"JPL, 2002.
\bibitem{ret65} M. Gibson, D. Poston, P. McClure, T. Godfroy, M. Briggs and J. Sanzi, "The Kilopower Reactor Using Stirling TechnologY (KRUSTY) Nuclear Ground Test Results and Lessons Learned," NASA Technical Report Server, 2017.
\bibitem{ret67} J. E. Werner, S. G. Johnson, C. C. Dwight and K. L. Lively, "Cost Comparison in 2015 Dollars for Radioisotope Power Systems—Cassini and Mars Science Laboratory," Idaho National Laboratory, Idaho Falls, 2016.
\bibitem{CasAttitude} C. Mittelsteadt, Cassini Attitude Control Operations - Guidelines Levied on Science To Extend Reaction Wheel Life, JPL Technical report Server, 2011. [Accessed 2020].
\bibitem{JunoCam} C. J. Hansen et. al., Junocam: Juno's Outreach Camera, Space Science Reviews, 2014. [Accessed 2020].
\bibitem{Pepssi} R. McNutt et. al., The Pluto Energetic Particle Spectrometer Science Investigation (PEPSSI) on the New Horizons Mission, Space Science Reviews, 2008. [Accessed 2020]. 
\bibitem{Reza} R. R. Karimi et. al., Mission Design for a Solar System Fast Escape to Interstellar Medium and Solar-Gravity Lens Focus, American Astronomical Society, 2020. [Accessed 2020].  DOI: 2020AGUFMSH0170009K
\bibitem{PSP} NASA/APL, "Parker Solar Probe: The Mission," jhuapl.com, 2019. [Online]. Available: http://parkersolarprobe.jhuapl.edu/The-Mission/index.php [Accessed 2020]
\bibitem{MALTO} Sims, Jon \& Finlayson, Paul \& Rinderle, Edward \& Vavrina, Matthew \& Kowalkowski, Theresa. (2006). Implementation of a Low-Thrust Trajectory Optimization Algorithm for Preliminary Design. Collection of Technical Papers - AIAA/AAS Astrodynamics Specialist Conference, 2006. 3. 10.2514/6.2006-6746. 
\bibitem{ANSIAIAA} “Standard: Mass Properties Control for Space Systems (ANSI/AIAA S-120A-2015(2019))”, in Standard: Mass Properties Control for Space Systems (ANSI/AIAA S-120A-2015(2019)), DOI:10.2514/4.103858.001.

\end{thebibliography}


\end{document}